\documentclass[11pt]{article}
\usepackage{amsfonts}
\usepackage{amsmath,amssymb,amscd}
\usepackage{stmaryrd}
\usepackage[dvips]{color}
\usepackage{fancybox}
\usepackage{cite}
\usepackage{bm}
\usepackage[all]{xy}
\usepackage{wrapfig}
\usepackage{wrapfig}   

\baselineskip=\normalbaselineskip

\def\Today{\ifcase\month\or January\or February\or March\or April\or May\or
 June\or July\or August\or September\or October\or November\or
December\fi\space\number\day, \number\year(\number\time)}

\def\TM{\widetilde{M}}

\def\condition#1
{\medskip\noindent{\bf Condition\ } #1 :\ }

\def\and{\quad{\rm and}\quad}

\def\p{\partial}

\def\mat#1{\left(\matrix{#1}\right)}

\def\IR{\relax{\rm I\kern-.18em R}}

\def\pr0#1#2#3{{\it Phys.\ Rev.} {\bf #1} (#2) #3}

\def\complex{{\mathchoice
{\setbox0=\hbox{$\displaystyle\rm C$}\hbox{\hbox to0pt
{\kern0.4\wd0\vrule height0.9\ht0\hss}\box0}}
{\setbox0=\hbox{$\textstyle\rm C$}\hbox{\hbox to0pt
{\kern0.4\wd0\vrule height0.9\ht0\hss}\box0}}
{\setbox0=\hbox{$\scriptstyle\rm C$}\hbox{\hbox to0pt
{\kern0.4\wd0\vrule height0.9\ht0\hss}\box0}}
{\setbox0=\hbox{$\scriptscriptstyle\rm C$}\hbox{\hbox to0pt
{\kern0.4\wd0\vrule height0.9\ht0\hss}\box0}}}}
\def\Co{{\mathchoice
{\setbox0=\hbox{$\displaystyle\rm C$}\hbox{\hbox to0pt
{\kern0.4\wd0\vrule height0.9\ht0\hss}\box0}}
{\setbox0=\hbox{$\textstyle\rm C$}\hbox{\hbox to0pt
{\kern0.4\wd0\vrule height0.9\ht0\hss}\box0}}
{\setbox0=\hbox{$\scriptstyle\rm C$}\hbox{\hbox to0pt
{\kern0.4\wd0\vrule height0.9\ht0\hss}\box0}}
{\setbox0=\hbox{$\scriptscriptstyle\rm C$}\hbox{\hbox to0pt
{\kern0.4\wd0\vrule height0.9\ht0\hss}\box0}}}}
\def\Rl{{\mathchoice
{\setbox0=\hbox{$\displaystyle\rm R$}\hbox{\hbox to0pt
{\kern0.4\wd0\vrule height0.9\ht0\hss}\box0}}
{\setbox0=\hbox{$\textstyle\rm R$}\hbox{\hbox to0pt
{\kern0.4\wd0\vrule height0.9\ht0\hss}\box0}}
{\setbox0=\hbox{$\scriptstyle\rm R$}\hbox{\hbox to0pt
{\kern0.4\wd0\vrule height0.9\ht0\hss}\box0}}
{\setbox0=\hbox{$\scriptscriptstyle\rm R$}\hbox{\hbox to0pt
{\kern0.4\wd0\vrule height0.9\ht0\hss}\box0}}}}

\begin{document} 

\numberwithin{equation}{section}
\renewcommand{\theequation}{\thesection.\arabic{equation}}
\def\natural{\mathbb{N}}
\def\mat#1{\matt[#1]}
\def\matt[#1,#2,#3,#4]{\left(%
\begin{array}{cc} #1 & #2 \\ #3 & #4 \end{array} \right)}
\def\hq{\hat{q}}
\def\hp{\hat{p}}
\def\hx{\hat{x}}
\def\hk{\hat{k}}
\def\hw{\hat{w}}
\def\hl{\hat{l}}

\def\bea#1\ena{\begin{align}#1\end{align}}
\def\bean#1\enan{\begin{align*}#1\end{align*}}
\def\nn{\nonumber\\}
\def\cL{{\cal L}}
\def\cE{{\cal E}}
\def\TM{TM\oplus T^*M}
\newcommand{\CouB}[2]{\left\llbracket #1,#2 \right\rrbracket}
\newcommand{\pair}[2]{\left\langle\, #1, #2\,\right\rangle}

\null \hfill Preprint TU-992, UTHEP-677  \\[3em]
\begin{center}
{\LARGE \bf{
Gravity theory on Poisson manifold with $R$-flux}}
\end{center}

\begin{center}
{T. Asakawa${}^\sharp$\footnote{
e-mail: asakawa@maebashi-it.ac.jp}, H. Muraki${}^{\flat}$\footnote{
e-mail: hmuraki@het.ph.tsukuba.ac.jp} and S. Watamura${}^{\dagger}$\footnote{
e-mail: watamura@tuhep.phys.tohoku.ac.jp}}\\[3em] 
${}^\sharp$
Department of Integrated Design Engineering,\\
Faculty of Engineering,\\
Maebashi Institute of Technology\\
Maebashi, 371-0816, Japan \\[1em]

${}^\flat$
Graduate School of Pure and Applied Sciences,\\
University of Tsukuba\\
Tsukuba, 305-8571, Japan \\[1em]

${}^\dagger$
Particle Theory and Cosmology Group, \\
Department of Physics, \\
Graduate School of Science, \\
Tohoku University \\
Aoba-ku, Sendai, 980-8578, Japan \\ [5ex]

\thispagestyle{empty}

\abstract{
\noindent
A novel gravity theory based on Poisson Generalized Geometry is investigated.
A gravity theory on a Poisson manifold equipped with a Riemannian metric
is constructed from a contravariant version
of the Levi-Civita connection, which is based on the Lie algebroid of a Poisson manifold.
Then, we show that in Poisson Generalized Geometry
the $R$-fluxes are consistently coupled with such a gravity.
An $R$-flux appears as a torsion of the corresponding connection in 
a similar way as an $H$-flux which appears as a torsion of the connection formulated
in the standard Generalized Geometry.
We give an analogue of the Einstein-Hilbert action coupled with an $R$-flux, 
and show that it is invariant under both $\beta$-diffeomorphisms and 
$\beta$-gauge transformations.
} 

\end{center}

\vskip 2cm

\eject

 \tableofcontents

\section{Introduction}

Poisson Generalized Geometry (PGG) \cite{Asakawa:2014kua,Asakawa:2015aia}
is a variant of Generalized Geometry (GG) \cite{Hitchin,Gualtieri1,Gualtieri2} in the sense that it
shares the same bundle $T M\oplus T^*M$, where $T M$ and $T^*M$ are tangent and cotangent
bundles  of a manifold $M$,
while the roles of vectors and $1$-forms are exchanged in PGG.
The bracket used in PGG defines a type of Courant algebroid,
which has its basis on a 
Lie algebroid $(T^*M)_\theta$ of a Poisson manifold $M$
\cite{Koszul,Bhaskara,KSPoisson}.

The symmetry of the Courant algebroid of PGG consists of
$\beta$-diffeomorphisms and $\beta$-transformations,
following the terminology introduced in 
\cite{Blumenhagen:2012nt,Andriot:2013xca}\footnote{
Strictly speaking, the $\beta$-diffeomorphism defined by the authors \cite{Blumenhagen:2012nt,Andriot:2013xca} 
is different from our definition. See \cite{Asakawa:2014kua}.}.
An $R$-flux, i.e. a totally antisymmetric tensor of type $(3,0)$, naturally arises in this framework
as an Abelian field strength
 associated with a twisting by local $\beta$-transformations \cite{Asakawa:2014kua}.
However, it is not yet clear that this flux can actually be interpreted as the 
``$R$-flux'' which is one of the non-geometric fluxes argued in physics literature
\cite{Dabholkar:2005ve,Shelton:2005cf,Hull:2004in}.
In order to clarify this point, 
we investigate in this paper a construction of gravity theory coupled with the $R$-flux of PGG,
since the non-geometric fluxes are considered mainly in the context of gravity theory
coupled with them.

In the framework of GG,
the underlying Riemannian geometry for a gravity theory coupled with an $H$-flux 
is investigated \cite{GualtieriPoisson,David2007,Hitchin:2010qz}.
It turns out that the resulting gravity is merely a lift of the usual general relativity, 
i.e. it is based on the Levi-Civita connection on the tangent bundle $TM$,
and the $H$-flux is incorporated as a torsion part added to this connection under the lift.
Eventually,  it turns out that it is the same as
the NS-NS sector of supergravity theory,
whose action is invariant under the symmetry of GG, that is,
diffeomorphisms and $B$-(field) gauge transformations.

In analogy with GG,
it is natural to expect that a gravity theory in PGG has similar structures to that in GG.
Namely, it has been suggested that the gravity theory would 
be a theory invariant  under $\beta$-diffeomorphisms and $\beta$-gauge transformations.
To show this, it is first required to formulate a 
``general relativity,"  which is invariant under $\beta$-diffeomorphisms on a Poisson manifold,
because no well-accepted theory of gravity of such kind has been known at least in the physics literature.
Following mathematical literature \cite{fernandes2000,Hawkins:2002rf,Boucetta:2011ofa1,Boucetta:2011ofa2,Boucetta:2011ofa3,Bruce:2015mla}, we construct such a gravity based on the Lie algebroid $(T^*M)_\theta$ of a Poisson manifold,
by replacing the Levi-Civita connection with its contravariant analogue.
We then extend this construction to that in PGG coupled to an $R$-flux
by applying the same strategy as used in GG \cite{GualtieriPoisson,David2007,Hitchin:2010qz}.
We show that an $R$-flux appears as a torsion part added to the 
contravariant analogue of the Levi-Civita connection under the lift,
in a similar manner as it is done in the case of an $H$-flux.
By introducing an appropriate integration measure,
we obtain an Einstein-Hilbert-like action for this gravity theory
invariant under both $\beta$-diffeomorphisms and $\beta$-gauge transformations. 
In this paper, we assume that the spacetime metric has Euclidean signature, 
though the extension to Lorentzian signature is straightforward.

The organization of this paper is as follows.
In section $2$, we study a Riemannian geometry that is compatible with a Poisson structure.
In section $3$, a construction of gravity theory based on Poisson generalized geometry
in the presence of an $R$-flux is investigated.
In section $4$, we summarize this paper.
Comparisons with other approaches 
are also discussed.
Since our constructions
are analogous to those of \cite{David2007,Hitchin:2010qz},
we give a short review on them in appendix A.
Appendix B is devoted to the computational details.

\section{Riemannian geometry on Poisson manifold}

The standard general relativity is based on Riemannian geometry, 
more precisely, it is based explicitly on a Riemannian metric on the tangent bundle $TM$ of a manifold $M$, 
and thus implicitly also on structures of the $TM$ as a Lie algebroid.
Then, if the underlying Lie algebroid $TM$ is replaced by a different Lie algebroid
$(T^*M)_\theta$ as introduced below,
the subsequent geometrical objects, such as connection, torsion, curvature etc., are also replaced.

More generally, in order to formulate a gravity theory associated with a Lie algebroid $A$, we need
the following materials:
\begin{enumerate}
\item
a Lie algebroid $A$ as an infinitesimal symmetry,
\item
a differential algebra $(\Gamma(\wedge^\bullet A^*), \wedge, d_A)$ as a differential calculus,
\item
a Riemannian metric on $A$,
\item
an affine $A$-connection $\nabla: \Gamma(A)\to \Gamma(A^*\otimes A)$ on the vector bundle $A$,
\item
a torsion and a Riemann curvature tensor of the affine connection $\nabla$,
\item
a notion of $A$-tensor fields, together with their transformation properties under  $A$,
\item
an invariant measure with respect to the symmetry  $A$,
\item
an Einstein-Hilbert like 
action.
\end{enumerate}
The first two objects in the list, $1$ and $2$, are well-known (see for example \cite{CdSWeinstein:1999}), 
and the objects in $3$-$5$ have already been studied in mathematical literature 
\cite{fernandes2000,Hawkins:2002rf,Boucetta:2011ofa1,Boucetta:2011ofa2,Boucetta:2011ofa3,Bruce:2015mla}.

In this section, we review the formalism for the case of $A=(T^*M)_\theta$, 
emphasizing a physicist's viewpoint
and give a study of $A$-tensor fields listed in point 6.
The remaining two points, $7$ and $8$, will be studied in the next section.

\subsection{Lie algebroid on a Poisson manifold}

Let $M$ be an $n$-dimensional Poisson manifold equipped with a Poisson bivector 
$\theta \in \Gamma(\wedge^2 T M)$.
Here, $\theta$ satisfies the Poisson condition $[\theta,\theta]_S=0$, 
where $[\cdot,\cdot]_S$ is the Schouten-Nijenhuis bracket.
A Lie algebroid of a Poisson manifold \cite{CdSWeinstein:1999} is defined 
by a triple $(T^*M, \, \theta, \, [\cdot,\cdot]_\theta)$,
where $T^*M$ is the cotangent bundle over $M$,
the anchor map $\theta :T^*M \to T M$ is defined by the Poisson bivector as
$\theta(\xi)=\bar{\iota}_\xi \theta$, for $\xi \in \Gamma(T^*M)$,
and the Lie bracket is defined by the Koszul bracket
\bea
[\xi,\eta]_\theta ={\cal L}_{\theta (\xi)}\eta-i_{\theta (\eta)}d\xi .
\ena
We denote this Lie algebroid $(T^*M,\,  \theta, \, [\cdot,\cdot]_\theta)$ as $(T^*M)_\theta$ for short.

We can define an exterior derivative as $d_\theta =[\theta, \cdot]_S $ 
on the space of polyvectors $ \Gamma(\wedge^\bullet TM)$,
which is nilpotent $d^2_\theta=0$ due to the Poisson condition.
For a function $f\in C^\infty (M)$, its action yields the Hamiltonian vector field:
\bea
d_\theta f =[\theta, f]_S =-\theta (df).
\ena
The actions of the ``Lie derivative'' $\bar{\cal L}_{\zeta}$ with $\zeta \in \Gamma(T^*M)$
on  a function $f$, a $1$-form $\xi$ and a vector field $X$ are given by\footnote{
We use the symbol $\bar{~}$ to distinguish the operations of differential calculus 
induced by 1-forms from the usual ones induced by vector fields.
For example, $\bar{\iota}_\xi$ denotes the ``interior product'' of a 1-form $\xi$.}
\bea
&\bar{\cal L}_{\zeta} f :=\bar{\iota}_\zeta d_\theta f, \nn
&\bar{\cal L}_{\zeta} \xi :=[\zeta,\xi]_\theta , \nn
&\bar{\cal L}_{\zeta} X :=(d_\theta \bar{\iota}_\zeta +\bar{\iota}_\zeta d_\theta) X,
\label{A Lie derivative}
\ena
respectively. These operations satisfy the following
Cartan relations on the space of polyvectors $\Gamma(\wedge^\bullet T M)$
\bea
\{ \bar{\iota}_\xi, \bar{\iota}_\eta\}=0, \quad
\{ d_\theta ,\bar{\iota}_\xi\}=\bar{\cal L}_\xi, \quad
[\bar{\cal L}_\xi, \bar{\iota}_\eta ]=\bar{\iota}_{[\xi,\eta]_\theta},\quad
[\bar{\cal L}_\xi, \bar{\cal L}_\eta ]=\bar{\cal L}_{[\xi,\eta]_\theta},\quad
[d_\theta,\bar{\cal L}_\xi ]=0.
\label{A Cartan}
\ena
With the use of these operations, 
a differential geometry based on the Lie algebroid $(T^*M)_\theta$ (instead of $TM$)
can be considered on a Poisson manifold.

\subsection{Contravariant Levi-Civita connection}

In this section we study 
Riemannian geometry compatible with a Poisson structure,
following \cite{fernandes2000,Hawkins:2002rf,Boucetta:2011ofa1,Boucetta:2011ofa2,Boucetta:2011ofa3}.

Let $M$ be a 
Riemannian manifold as well as a 
Poisson manifold with local coordinates $\{x^i\}$,
and $G$ be a ``Riemannian metric'' on the cotangent bundle $T^*M$, i.e. $G$ defines
a positive-definite bilinear form on the fiber of $T^*M$.
Using the coordinate basis of $1$-forms $\{dx^i\}$, the bilinear form  is locally written as
\bea
G(\xi,\eta )=G^{ij}\xi_i \eta_j,
\ena
for any $1$-forms $\xi$, $\eta \in \Gamma(T^*M)$, where $G^{ij}:=G(dx^i,dx^j)$.
Its matrix inverse $G_{ij}$, 
i.e. $G_{ij}G^{jk}=\delta_i^k$, defines a metric on the 
tangent bundle $TM$, so that $G^{-1}$ is an ordinary Riemannian metric.

\paragraph{Contravariant connection}
Since we regard the Lie algebroid $(T^*M)_\theta$ as a 
fundamental object  of the geometry of our interest, 
the notion of a connection is changed as follows \cite{fernandes2000}.

Let $E$ be a vector bundle over $M$.
A contravariant connection on $E$ is a linear map 
$\bar{\nabla}:\Gamma(E) \to\Gamma(TM \otimes E)$
satisfying
\bea
	\bar\nabla (fs)
	= d_\theta f \otimes s + f \bar\nabla s,\label{condition of contra connecttion}
\ena
for any section $s \in \Gamma(E)$ and function $f\in C^\infty (M)$.
This is a generalization of the exterior derivative $d_\theta$ so as to act on a vector bundle $E$.
We refer to $\bar\nabla s$ as the contravariant derivative of $s$.

\paragraph{Contravariant affine connection}

In particular, a contravariant connection on the cotangent bundle $E=T^*M$ should be called 
a contravariant affine connection.
This is also understood
 as a bilinear map $\bar\nabla :\Gamma(T^*M) \times \Gamma(T^*M)\to\Gamma(T^*M)$;
$(\xi,\eta) \mapsto \bar\nabla_\xi \eta$, 
such that, for any $1$-forms $\xi$, $\eta$ and function $f$, 
\bea
\bar\nabla_{f\xi} \eta =f \bar\nabla_\xi \eta, 
~~\bar\nabla_\xi (f\eta)=(\bar\cL_{\xi} f)\eta+f \bar\nabla_{\xi} \eta . \label{affine}
\ena
The point is that 
the argument $\xi$ in the directional derivative $\bar\nabla_\xi$ is given by a $1$-form.
The connection coefficients with respect to the coordinate basis $\{dx^i\}$ are defined through
\bea
	\bar\nabla_{dx^i}dx^j = \bar{\Gamma}^{ij}_k dx^k. \label{25}
\ena
Hence, together with \eqref{affine}, we have
\bea
	\bar\nabla_{\xi} \eta
	= \xi_i (\theta^{ij}\p_j \eta_k +\bar{\Gamma}^{ij}_k \eta_j )dx^k ,\label{nablaetaxi local}
\ena
for $\xi=\xi_idx^i$ and $\eta=\eta_idx^i$ in local coordinates $\{x^i\}$.
It is worth comparing the formulae above with those obtained in the tangent bundle $TM$
by using an ordinary affine connection $\nabla_X$ and connection coefficients
$\nabla_{i}\p_j = \Gamma^k_{ij} \p_k$ with respect to the basis vectors $\{\p_i\}$.

\paragraph{Contravariant torsion}

The torsion of a contravariant affine connection $\bar \nabla$ is defined 
by
\bea
\bar{T}(\xi,\eta)=\bar\nabla_{\xi} \eta -\bar\nabla_{\eta} \xi - [\xi,\eta]_\theta,\label{2.15}
\ena
for any $1$-forms $\xi$ and $\eta$. 
Since it is manifestly skew-symmetric and satisfies 
\bea
\bar{T}(f\xi,\eta)
=f\bar{T}(\xi,\eta)
=\bar{T}(\xi,f\eta),
 \label{multi-linearity}
\ena
$\bar{T}$ is a tensor in $\Gamma(\wedge^2 TM \otimes T^*M)$.
We call $\bar T$ the contravariant torsion.

\paragraph{Contravariant Levi-Civita connection}
In the usual Riemannian geometry, 
the Levi-Civita connection $\nabla$ on the tangent bundle $TM$ is
a unique torsion-free metric connection.
In an analogous way, a contravariant affine connection $\bar{\nabla}$ is
called a contravariant Levi-Civita connection, 
if it is compatible with the metric and torsion free:
\bea
&\bar\cL_{\xi} G(\eta,\zeta)=G (\bar{\nabla}_\xi \eta, \zeta) +G (\eta, \bar{\nabla}_\xi \zeta) ,\label{28}\\
&\bar{T}(\xi,\eta)=0,\label{torsion-free condition}
\ena
for arbitrary $1$-forms $\xi$, $\eta$ and $\zeta$.

We show that the contravariant Levi-Civita connection on $T^*M$ is uniquely
specified by the Koszul formula\footnote{For a proof, see Appendix \ref{compdet}.}
\cite{Boucetta:2011ofa1,Boucetta:2011ofa2,Boucetta:2011ofa3} 
\bea
2G (\bar{\nabla}_\xi \eta, \zeta) 
&=\theta (\xi)\cdot G(\eta,\zeta)+\theta (\eta) \cdot G(\xi,\zeta)-\theta (\zeta) \cdot G(\xi,\eta)\nn
~~&+G([\zeta,\xi]_\theta,\eta)+G([\zeta,\eta]_\theta,\xi)+G([\xi,\eta]_\theta,\zeta). \label{Koszulforml}
\ena
Through this formula, the connection is
determined only by the Riemannian metric $G$ on $T^*M$ and the 
information of the Lie algebroid $(T^*M)_\theta$ (i.e. the anchor map and the Koszul bracket). 

In local coordinates, with the use of (\ref{25}), the formula \eqref{Koszulforml} yields
\bea
2G^{lk}\bar{\Gamma}^{ij}_l 
=\theta^{il} \p_l G^{jk}+\theta^{jl} \p_l G^{ik}-\theta^{kl} \p_l G^{ij}
+G^{lj}\p_l \theta^{ki}+G^{li}\p_l \theta^{kj}+G^{lk}\p_l \theta^{ij}.  \label{Koszulforml^}
\ena
where 
\bea
& G(\bar{\nabla}_{dx^i} dx^j, dx^k)
=G(\bar{\Gamma}^{ij}_l dx^l, dx^k)=G^{lk}\bar{\Gamma}^{ij}_l ,\nn
& \theta(dx^i)\cdot G(dx^j,dx^k)=\theta^{il} \p_l G^{jk},\nn
& G([dx^k,dx^i]_\theta ,dx^j)=G(\p_l \theta^{ki}dx^l ,dx^j)=G^{lj}\p_l \theta^{ki}.
\ena
By acting $G_{k'k}$ on both sides of \eqref{Koszulforml^} and rewriting $k'$ to $k$, it is reduced to 
\bea
\bar{\Gamma}^{ij}_k
&=\frac{1}{2}G_{nk}\left(\theta^{il} \p_l G^{jn}+\theta^{jl} \p_l G^{in}-\theta^{nl} \p_l G^{ij}
+G^{lj}\p_l \theta^{ni}+G^{li}\p_l \theta^{nj}+G^{ln}\p_l \theta^{ij}\right).\label{220}
\ena
In contrast to the usual Christoffel symbol $\Gamma_{ij}^k$ that is specified only by the metric,
the contravariant Christoffel symbol 
$\bar{\Gamma}^{ij}_k$ \eqref{220} is determined by both the metric and the Poisson tensor.

Note that the condition of compatibility with the metric (\ref{28}) written in local coordinates implies
\bea
	\theta^{kl}\p_l G^{ij}-\bar{\Gamma}^{ki}_lG^{lj}-\bar{\Gamma}^{kj}_l G^{il}
	 =  0. \label{g1}
\ena
This equation will be rewritten in a more familiar expression 
$\bar \nabla_{dx^k} G^{ij}=0$ in \S \ref{sec:tensor}. 
The torsion-free condition (\ref{torsion-free condition}) in terms of components
indicates 
\bea
	\bar{T}^{ij}_k &=\bar{\Gamma}^{ij}_k-\bar{\Gamma}^{ji}_k -\p_k \theta^{ij}
	= 0. \label{Ttheta}
\ena 
Thus, it implies that the anti-symmetric part of the contravariant Christoffel symbol 
should not vanish
\bea
	\bar{\Gamma}^{ij}_k-\bar{\Gamma}^{ji}_k = \p_k \theta^{ij}.	\label{218}
\ena
This is a consequence of the non-vanishing Lie bracket
$[dx^i,dx^j]_\theta=\p_k \theta^{ij}dx^k$ of the coordinate basis, which
is a significant difference from the ordinary Christoffel symbol $\Gamma_{ij}^k$
using the standard Lie algebroid $TM$, where $[\p_i,\p_j]_{T M}=0$%
\footnote{The anti-symmetric part does not vanish 
in general for a non-holonomic frame $e_a=e_a^i \p_i$ of $TM$.}.

\paragraph{Contravariant curvature}

The curvature of a contravariant affine connection $\bar \nabla$ is defined by
\bea
	\bar{R}(\xi,\eta)\zeta=
	(\bar{\nabla}_\xi \bar{\nabla}_\eta -\bar{\nabla}_\eta \bar{\nabla}_\xi 
		- \bar \nabla_{[\xi,\eta]_\theta} )\zeta,
\ena
where $\xi,\eta$ and $\zeta\in\Gamma(T^*M)$.
It is easily shown that $\bar{R}$ is tensorial since it satisfies
\bea
	\bar{R}(f\xi,g\eta)(h\zeta)=fgh\bar{R}(\xi,\eta)\zeta,
\ena
for any function $f,g$ and $h$.
Thus it defines a map $\bar{R}:\Gamma(T^*M)\otimes \Gamma(T^*M) \to {\rm End}(T^*M)$ 
and we will refer to it as contravariant curvature (Riemann) tensor hereafter.

For this curvature, together with the torsion $\bar{T}$, 
the following Bianchi identities hold\footnote{See Appendix \ref{compdet},
for these derivations.}: 
\bea
	&\mathfrak{S}\{\bar{R}(\xi,\eta) \zeta\} 
	=\mathfrak{S}\{(\bar{\nabla}_{\zeta}\bar{T})(\xi,\eta)
	+\bar{T}(\bar{T}(\xi,\eta),\zeta)\},\label{bi1}\\
	&\mathfrak{S}\{(\bar{\nabla}_\zeta \bar{R})(\xi,\eta)
	+\bar{R}(\bar{T}(\xi,\eta),\zeta)\}=0,\label{bi2}
\ena
where $\mathfrak{S}$ denotes the cyclic sum over $\xi$, $\eta$, and $\zeta$, e.g. 
$\mathfrak{S}\{\bar{R}(\xi,\eta) \zeta \}=\bar{R}(\xi,\eta) \zeta+\bar{R}(\eta,\zeta) \xi+\bar{R}(\zeta,\xi) \eta$.

In local coordinates, the components of the curvature are read off from 
$\bar{R}(dx^i,dx^j)dx^k=:\bar{R}^{kij}_{l} dx^l$ and then we have
\bea
	 \bar{R}^{kij}_{l} 
	&= \theta^{im} \p_m \bar{\Gamma}^{jk}_l 
		- \theta^{jm} \p_m \bar{\Gamma}^{ik}_l -\p_n \theta^{ij} \bar{\Gamma}^{nk}_l 
		+ \bar{\Gamma}^{jk}_m \bar{\Gamma}^{im}_l 
		 - \bar{\Gamma}^{ik}_m \bar{\Gamma}^{jm}_l  ,\label{contra Riem}
\ena
where $\bar{\Gamma}^{ij}_k$  are connection coefficients, and are in particular
given by (\ref{220}) for the case of the contravariant Levi-Civita connection.

A contravariant analogue of the Ricci tensor is defined
by contracting an upper index with a lower one:
\bea
	\bar{R}^{kj}:=&\bar{R}^{klj}_l,\label{contra Ricci}
\ena
and an analogue of the scalar curvature
is defined by taking the contraction between the metric  $G$ and the analogue of the Ricci tensor
\bea
	\bar{R}&:=G_{kj}\bar{R}^{kj}.\label{contra Ricci scalar}
\ena

\subsection{Tensor calculus \label{sec:tensor}}

In a geometry with an ordinary affine connection $\nabla$ of the tangent bundle $T M$, 
the covariant derivative  has a canonical extension
that can act on any type of tensor field.
A tensor field $T$ of type $(r,s)$ is a section of the tensor bundle 
$(TM)^{\otimes r}\otimes (T^*M)^{\otimes s}$, which is also characterized 
by the transformation rules under diffeomorphisms\footnote
{As a transformation rule, we may either use 
a general coordinate transformation (passive) 
or a diffeomorphism (active). See a nice lecture \cite{Carroll:1997ar} on this point.}.
The covariant derivative $\nabla$ is by definition a covariant object, in the sense that 
$\nabla_X T$ is an $(r,s)$-tensor field, if so $T$ is.
These facts serve as the basis of the tensor calculus.

In this subsection, we give similar arguments
in the case of a contravariant affine connection.

\paragraph{Tensor fields}
We first need to introduce an appropriate notion of tensor fields.
Fortunately, an ordinary tensor field $T$ is automatically a tensor field for the contravariant case.
This is due to the fact that the notion of tensor bundles is unchanged.
Thus, for example, the contraction 
of a vector field $X$ and a $1$-form $\eta$, $\bar\iota_X \eta=X^j \eta_j$, is a scalar field, as usual.
However, the infinitesimal 
transformation characterizing a tensor field is 
now given by the action of the Lie derivative $\bar {\cal L}_\zeta$
 generated by any $1$-form $\zeta$.
We call them $\beta$-diffeomorphisms according to \cite{Asakawa:2014kua}.

By rewriting (\ref{A Lie derivative}) in terms of local coordinates, 
we find that a vector field $X$ and a $1$-form $\eta$ transform under a 
$\beta$-diffeomorphism generated by $\zeta$, (here we denote $\p^i=\theta^{ij}\p_j$), 
respectively, as
\bea
&\bar {\cal L}_\zeta X^i =\zeta_k \p^k X^i +\p^i \zeta_l X^l 
+\p_l \theta^{ik}\zeta_k X^l, \nn
&\bar {\cal L}_\zeta \eta_j =\zeta_k \p^k \eta_j -\p^l \zeta_j \eta_l 
-\p_j \theta^{lk}\zeta_k \eta_l.\label{beta 1-form law}
\ena
These formulae are also understood in more conventional way as follows.

The $\beta$-diffeomorphism acts on a coordinate function $x^i$ as
\bea
\bar {\cal L}_\zeta x^i=-\bar \iota_\zeta \theta (dx^i) 
=\theta(\zeta,dx^i) =\theta^{ki}\zeta_k,
\ena
so that the first term 
of each equation in \eqref{beta 1-form law} 
comes from the shift of the argument $x^i\to x^i -\theta^{ki}\zeta_k$.
It acts also on the coordinate basis $dx^i$ of $T^*M$ as
\bea
\bar {\cal L}_\zeta dx^i =[\zeta, dx^i]_\theta =([\zeta, dx^i]_\theta)_j dx^j ,
\ena
so that the matrix $M(\zeta)$  of the change of basis in $dx^i \to dx^i + M(\zeta)^i_j dx^j$ is given by
\bea
M(\zeta)^i_j =([\zeta, dx^i]_\theta)_j 
=-(\p^i \zeta_j +\p_j \theta^{il} \zeta_l ).
\ena
The remaining terms in \eqref{beta 1-form law} are exactly the result of this change of basis,
since they are rewritten in terms of $M(\zeta)$ as
\bea
&\bar {\cal L}_\zeta X^i =\zeta_k \p^k X^i - M(\zeta)^i_l X^l , \nn
&\bar {\cal L}_\zeta \eta_j =\zeta_k \p^k \eta_j + M(\zeta)^l_j \eta_l .\label{beta 1-form law 2}
\ena
Thus, there are two familiar contributions to the change of vectors and $1$-forms under 
an infinitesimal $\beta$-diffeomorphism \eqref{beta 1-form law}: One is
the shift of the argument and the other is the tensorial factor, 
as in the case of an ordinary infinitesimal diffeomorphisms%
\footnote{
Note that in the ordinary case, the shift $x^i\to x^i+\epsilon^i $ and the change of basis 
$dx^i\to dx^i+\p_j \epsilon^i dx^j$ are essentially the same because of $[{\cal L}_\epsilon ,d]=0$,
while in our case they are not due to $[\bar{\cal L}_\zeta ,d]\neq0$}.

With this observation, an extension to characterize tensor fields is canonically achieved:
An $(r,s)$-tensor field $T$ transforms under the above $\beta$-diffeomorphism as
\bea
\bar {\cal L}_\zeta T^{i_1 \cdots i_r}_{j_1 \cdots j_s}
&=\zeta_k\p^k T^{i_1 \cdots i_r}_{j_1 \cdots j_s}
-\sum_{p=1}^r  M(\zeta)^{i_p}_l T^{i_1 \cdots l \cdots i_r}_{j_1 \cdots j_s}
+\sum_{q=1}^s  M(\zeta)^l_{j_q} T^{i_1 \cdots i_r}_{j_1 \cdots l \cdots j_s}.
\label{beta tensor law}
\ena
Note that the roles of upper and lower indices are switched and there appear relative signs, 
as compared to the ordinary transformation law of an $(r,s)$-tensor field $T$ under diffeomorphism.
In spite of such differences, 
the notion of contraction between
upper and lower indices of tensors still works,
since only the indices that are not contracted determine the transformation properties
under $\beta$-diffeomorphisms.
For instance, $X^j \eta_j$ transforms as a scalar field
\bea
\bar {\cal L}_\zeta (X^j\eta_j) &=(\zeta_k \p^k X^j -M(\zeta)^j_l X^l) \eta_j 
+X^j (\zeta_k \p^k \eta_j + M(\zeta)^l_j \eta_l) \nn
&=\zeta_k \p^k (X^j\eta_j),
\ena
and $G^{ij}\eta_j$ transforms as a vector field, and so on.

\paragraph{Contravariant derivative}
The actions of the contravariant derivative of an affine connection $\bar{\nabla}_\zeta$ 
on the components of a vector field, $X^i$, and a $1$-form, $\eta_j$, are written by 
\bea
&\bar{\nabla}_\zeta \eta_j 
=\zeta_k \bar{\nabla}_{dx^k} \eta_j 
=\zeta_k \left( \p^k \eta_j 
+\bar{\Gamma}^{kl}_{j} \eta_l \right),\nn
&\bar{\nabla}_\zeta X^i 
=\zeta_k \bar{\nabla}_{dx^k} X^i
=\zeta_k \left( \p^k X^i 
-\bar{\Gamma}^{ki}_l X^l \right), \label{contra 1-form law}
\ena
respectively.
The former comes from the connection on $T^*M$, given in (\ref{nablaetaxi local}).
The latter comes from the induced connection on the dual bundle $TM$, which is 
determined by the compatibility of the contravariant derivative with the contraction:
\bea
	\bar{\cL}_{\zeta} (X^j \eta_j) =(\bar{\nabla}_{\zeta}X^j)\eta_j + X^j (\bar{\nabla}_{\zeta} \eta_j).
\ena

From the formulae \eqref{contra 1-form law}, the action of the contravariant derivative
 is straightforwardly extended to an $(r,s)$-tensor field $T$\footnote{For a detailed analysis, 
 see appendix \ref{compdet}.}:
\bea
\bar{\nabla}_\zeta T^{i_1 \cdots i_r}_{j_1 \cdots j_s}
&=\zeta_k \left( \p^k T^{i_1 \cdots i_r}_{j_1 \cdots j_s}
-\sum_{p=1}^r \bar{\Gamma}^{ki_p}_l 
T^{i_1 \cdots l \cdots i_r}_{j_1 \cdots j_s}
+\sum_{q=1}^s \bar{\Gamma}^{kl}_{j_q} 
 T^{i_1 \cdots i_r}_{j_1 \cdots l \cdots j_s} \right).
\label{contra dedivative law}
\ena
Note again that the indices' structures are flipped
and the relative signs appear as compared to the case of the usual covariant derivative.
In the case of the contravariant Levi-Civita connection, 
the connection coefficients are specified by (\ref{220}).

By definition, $\bar \nabla_\zeta \eta_j$ should transform as components of a $1$-form.
In order to achieve this, the connection coefficients are found to transform 
under the $\beta$-diffeomorphism not as a tensor but as
\bea
\delta_\zeta \bar \Gamma^{il}_j 
&=\zeta_k \p^k \bar \Gamma^{il}_j
+M(\zeta)^k_j \bar \Gamma^{il}_k -M(\zeta)^i_k \bar \Gamma^{kl}_j -M(\zeta)^l_k \bar \Gamma^{ik}_j
-\p^i M(\zeta)^l_j .
\label{trf law of bar Gamma}
\ena
For details, see Appendix \ref{compdet}.
The first term comes form the shift of the argument, 
the three terms in the middle are the tensor factors,
and the last term is a non-tensorial factor, peculiar to the connection coefficients.
By using this property,
it is possible to show that the contravariant derivative of any other tensor
$\bar{\nabla}_\zeta T^{i_1 \cdots i_r}_{j_1 \cdots j_s}$ transforms
as an $(r,s)$-tensor again, see Appendix \ref{compdet}.

\paragraph{Contravariant derivative of $G$ and $\theta$}

There are two characteristic tensor fields $G^{ij}$ and $\theta^{ij}$ in the present formulation.
It follows from (\ref{contra dedivative law}) that the contravariant derivative of  the metric reads
\bea
	\bar{\nabla}_{dx^i} G^{mn}
	&= \theta^{il}\p_l G^{mn} 
	- \bar{\Gamma}^{im}_{l} G^{ln} 
		 - \bar{\Gamma}^{in}_{l} G^{ml}.
\ena
As mentioned in (\ref{g1}), it automatically vanishes, i.e.
$\bar{\nabla}_{dx^i} G^{mn}=0$, for the case of the contravariant Levi-Civita connection (\ref{220}),
owing to the metric-compatibility condition.

The contravariant derivative of the Poisson tensor is
\bea
	\bar{\nabla}_{dx^i} \theta^{mn}
	&= \theta^{il}\p_l \theta^{mn} 
	- \bar{\Gamma}^{im}_{l} \theta^{ln} 
		 - \bar{\Gamma}^{in}_{l} \theta^{ml}, \label{243}
\ena
which does not vanish even for the case of the contravariant Levi-Civita connection.
However,
permuting the indices of \eqref{243} cyclically and taking their sum, we see that
\bea
	&\bar\nabla_{dx^k} \theta^{ij}
	+ \bar\nabla_{dx^i} \theta^{jk}+ \bar\nabla_{dx^j} \theta^{ki}=0,\label{contra Poisson condition}
\ena
with the use of the Poisson condition $\theta^{m[i}\p_m\theta^{jk]}=0$ and \eqref{218}.
This result is not an accidental one but is a consequence of the torsion-free condition, as seen below.

In general, the exterior derivative $d_\theta$ acting on any polyvector can 
be written in terms of a contravariant affine connection $\bar{\nabla}$, if it is torsion-free\footnote{
There is a similar statement for the case of de Rham's exterior differential $d$ and 
an affine connection $\nabla$ acting on differential forms. }.
For example, for a vector field $X$, we find
\bea
d_\theta X (\xi,\eta) 
&=\theta (\xi) \cdot X (\eta) -\theta (\eta) \cdot X (\xi) -X([\xi,\eta]_\theta ) \nn
&=(\bar{\nabla}_\xi X) (\eta) +X (\bar{\nabla}_\xi \eta) 
-(\bar{\nabla}_\eta X) (\xi) - X (\bar{\nabla}_\eta \xi) -X([\xi,\eta]_\theta ) \nn
&=(\bar{\nabla}_\xi X) (\eta) -(\bar{\nabla}_\eta X) (\xi) 
+X(\bar{\nabla}_\xi \eta -\bar{\nabla}_\eta \xi -[\xi,\eta]_\theta ) \nn
&=(\bar{\nabla}_\xi X) (\eta) -(\bar{\nabla}_\eta X) (\xi),
\ena 
where in the first equality we used the definition of $d_\theta$,
and in the last equality we made use of the torsion-free condition.
Similarly, for $\theta \in \Gamma(\wedge^2 TM)$, we have 
\bea
d_\theta \theta (\xi,\eta,\zeta) 
&=\theta (\xi) \cdot \theta (\eta,\zeta)  +\theta (\eta) \cdot \theta (\zeta,\xi)  
+\theta (\zeta) \cdot \theta (\xi,\eta) 
-\theta ([\xi,\eta]_\theta,\zeta ) -\theta ([\eta,\zeta]_\theta,\xi )
-\theta ([\zeta,\xi]_\theta,\eta )  \nn
&= (\bar{\nabla}_\xi \theta) (\eta,\zeta)  +(\bar{\nabla}_\eta \theta) (\zeta,\xi)  
+(\bar{\nabla}_\zeta \theta) (\xi,\eta) \nn
&~~+\theta (\bar{\nabla}_\xi \eta -\bar{\nabla}_\eta \xi -[\xi,\eta]_\theta,\zeta ) 
+\theta (\bar{\nabla}_\eta \zeta -\bar{\nabla}_\zeta \eta -[\eta,\zeta]_\theta,\xi )
+\theta (\bar{\nabla}_\zeta \xi -\bar{\nabla}_\xi \zeta -[\zeta,\xi]_\theta,\eta ) \nn
&= (\bar{\nabla}_\xi \theta) (\eta,\zeta)  +(\bar{\nabla}_\eta \theta) (\zeta,\xi)  
+(\bar{\nabla}_\zeta \theta) (\xi,\eta).
\ena 
This means, in components, that 
\bea
d_\theta \theta 
&=\frac{1}{3!} (\bar{\nabla}_{dx^k} \theta^{ij} 
+\bar{\nabla}_{dx^i} \theta^{jk}+\bar{\nabla}_{dx^j} \theta^{ki})
\p_k\wedge \p_i \wedge \p_j
\ena
Thus, the Poisson condition $[\theta,\theta]_S=0$,  in terms of components, $\theta^{m[i}\p_m\theta^{jk]}=0$,
is equivalent to (\ref{contra Poisson condition}) as mentioned above.

\subsection{Ordinary Levi-Civita connection as contravariant affine connection}\label{cova-contra}

Before closing this section, here
we give a possible interpretation of
the contravariant 
Christoffel symbol $\bar{\Gamma}^{ij}_k$ (\ref{220}), 
by rewriting it in terms of the ordinary 
Christoffel symbol $\Gamma^k_{ij}$.
This rewriting is possible, since the derivative of the metric $G$ can be written as
\bea
	\p_m G^{ij}= - \Gamma^i_{ml}G^{lj} - \Gamma^j_{ml}G^{il},\label{pG}
\ena
using the ordinary Christoffel symbol $\Gamma^k_{ij}$.
Substituting (\ref{pG}) into the formula of the contravariant Christoffel symbol (\ref{220}), we find
\bea 
	\bar{\Gamma}^{ij}_k
	=\Gamma^j_{mk} \theta^{mi}    +K^{ij}_k.\label{covariant-contravariant}
\ena
Here, we introduced a tensor $K^{ij}_k$ defined by
\bea
& {K_k}^{ij}=G_{kn}K^{nij},
~~K^{kij} =\frac{1}{2}\left(\nabla^k \theta^{ij} -\nabla^i \theta^{jk}
+\nabla^j \theta^{ki}\right) ,
\ena
where $\nabla$ denotes the ordinary Levi-Civita connection and
the raising and lowering of indices is done by the metric $G$ as usual, 
e.g. $\nabla^i=G^{ij}\nabla_j$.

In eq. (\ref{covariant-contravariant}), 
$\Gamma^j_{km}$ is the ordinary Christoffel symbol associated with the
directional derivative $\nabla_X $ along a vector field $X\in \Gamma(TM)$.
However, if $X=\theta (\xi) $ for some $1$-form $\zeta \in \Gamma(T^*M)$, then
$\nabla_X=\nabla_{\theta (\xi)}$ 
might also be regarded as a contravariant derivative in the $\xi$-direction.
The combination $\Gamma^j_{km}\theta^{mi}$ appearing in (\ref{covariant-contravariant})
is indeed understood in this way.
That is, $\nabla_{\theta (\cdot)}$ 
defines another contravariant connection \cite{Hawkins:2002rf}, 
and the formula (\ref{covariant-contravariant})
indicates the difference between the two contravariant affine connections 
\bea
\nabla_{\theta (\cdot)} =\bar\nabla +K.
\ena
For example, we have 
\bea
& \nabla_{\theta (\xi)} X=\bar{\nabla}_{\xi} X +K(\xi,X), \nn
& \nabla_{\theta (\xi)} \eta =\bar{\nabla}_{\xi} \eta +K(\xi,\eta).
\ena

In general, the difference of any two contravariant affine connections 
should be an endmorphism-valued vector field.
Thus ${K_k}^{ij}$ should be a component of the tensor $K \in  \Gamma(T^*M\otimes TM \otimes TM)$.
In particular, since $\bar{\nabla}$ is the contravariant Levi-Civita connection, 
the difference $K$ should be referred to as a (contravariant version of the) contorsion tensor.
In fact, we can compute the contravariant torsion (\ref{2.15}) associated with 
the contravariant affine connection $\nabla_{\theta (\cdot)}$ as
\bea
{T}(\xi,\eta) 
&=\nabla_{\theta (\xi)}\eta -\nabla_{\theta (\eta)}\xi -[\xi,\eta]_\theta \nn
&=\bar{\nabla}_{\xi}\eta -\bar{\nabla}_{\eta}\xi -[\xi,\eta]_\theta + K(\xi,\eta) -K(\eta,\xi)\nn
&=K(\xi,\eta) -K(\eta,\xi),
\ena
which is in components 
\bea
{T}_k^{ij}={K_k}^{ij}-{K_k}^{ji}=\nabla_k \theta^{ij}.
\ena
Thus, the relation between the contorsion tensor $K$ and the torsion tensor ${T}$ is obtained as usual.

In constructing a gravity theory in the present framework using contravariant connections,
the theory based on the contravariant Levi-Civita $\bar{\nabla}$
would correspond to the contravariant analogue of Einstein's general relativity,
while the theory based on the metric affine $\nabla_{\theta (\cdot)}$ 
corresponds to a contravariant version of the 
Einstein-Cartan theory of gravity, equipped with a torsion.

On the other hand, 
the relation \eqref{covariant-contravariant} could 
provide another description of our current framework
from the viewpoint of the ordinary framework using covariant connections:
a deformation of general relativity by a matter field $\theta^{ij}$.
Let us rewrite the contravariant curvature tensors in the previous subsection by using 
(\ref{covariant-contravariant}).
After some lengthy computation, see appendix \ref{sriempois},
we obtain the contravariant Riemann curvature as
\bea
	&\bar{R}^{kij}_{l}
	= \theta^{im} \theta^{nj} {\sf R}^k_{lmn}
		-  (\nabla_n \theta^{ij}) K^{nk}_l
 		+ \theta^{nj} \nabla_n K^{ik}_l 
		 -\theta^{ni} \nabla_n K^{jk}_l
		+K^{jk}_m  K^{im}_l 
		-  K^{ik}_mK^{jm}_l  , \label{riempois}
\ena
where ${\sf R}^k_{lmn}$ denotes the ordinary Riemann curvature tensor made out of the metric $G$.
By using the expression \eqref{riempois}, the contravariant Ricci tensor and the scalar curvature are also written as 
\bea
	&\bar{R}^{kj}
  =\theta^{lm} \theta^{nj} {\sf R}^k_{lmn}
		-  (\nabla_n \theta^{lj}) K^{nk}_l
 		+ \theta^{nj} \nabla_n \nabla_l \theta^{lk} 
		 -\theta^{nl} \nabla_n K^{jk}_l
		+K^{jk}_m  \nabla_l \theta^{lm} 
		-   K^{lk}_m K^{jm}_l   ,\nn
	&\bar{R}
	= \theta^{lm} \theta^{nj} {\sf R}_{jlmn}
 		+ 2\theta_{nm} \nabla^n \nabla_l \theta^{lm} 
				- \nabla^n \theta_{nm}  \nabla_l \theta^{lm} .	
\ena
Thus, the appearance of these expressions in an ordinary gravity theory suggests 
the existence of a
dual description based on a contravariant connection.

As a particular case, when the contravariant contorsion tensor vanishes, $K^{ijk}=0$,
the contravariant curvatures reduces to the Riemann curvature, 
up to a multiplications by $\theta$.
This happens if $\nabla^n\theta^{ij}=0$, i.e. the Poisson tensor is covariantly constant.
This is because the requirement $K^{ijk}=0$ implies that
\bea
	\nabla^n \theta^{ij} 
		 =   \nabla^i \theta^{jn}
		 +   \nabla^j \theta^{in} .
\ena
By noticing that the left-hand side is skew-symmetric under the interchange of indices $\{i,j\}$,
whereas the right-hand side is symmetric, it follows that $\nabla^n\theta^{ij}=0$.
This class of geometry is studied in \cite{Boucetta:2011ofa1,Boucetta:2011ofa2,Boucetta:2011ofa3} under the name 
of a Riemann-Poisson 
manifold\footnote{In the literatures,
the author's definition ${\cal D}\pi=0$ is equivalent to $\nabla^n\theta^{ij}=0$.}.
This is realized, for example, when the manifold is equipped with a K\"ahler structure,
where the Poisson tensor is induced by the corresponding symplectic form.

\section{Gravity theory based on Poisson Generalized Geometry}

In this section, we address the construction of a gravity theory on a Poisson manifold
 in the framework of the Poisson Generalized Geometry (PGG) \cite{Asakawa:2014kua}, where
the Courant algebroid $(TM)_0 \oplus (T^*M)_\theta$ is the main object under consideration.
Applying Hitchin's constructions of a Generalized Riemannian geometry
for the standard Courant algebroid $TM \oplus T^*M$ \cite{Hitchin:2010qz} to our case,
we show that the connection, the curvature, etc. can be constructed in an analogous way.
In particular, the $R$-flux formulated in PGG \cite{Asakawa:2014kua} appears as a torsion, 
similarly as the $H$-flux does.
After investigating an invariant integration measure, we will give an analogue of the Einstein-Hilbert action
on a Poisson manifold.
For reference purposes, we give a brief review on related issues of \cite{Hitchin:2010qz}
in appendix \ref{RGGG}.

\subsection{Courant algebroid $(T M)_0\oplus (T^*M)_\theta$}

Consider a vector bundle $T M\oplus T^*M$ equipped with a canonical inner product defined
for any vector fields $X$, $Y$ and $1$-forms $\xi$, $\eta$ with 
$X+\xi, Y+\eta \in \Gamma(T M\oplus T^*M)$ as
\bea
\langle X+\xi,Y+\eta\rangle ={\frac{1}{2}}(\bar{\iota}_\xi Y+\bar{\iota}_\eta X ),	\label{Sasa inner}
\ena
with an anchor map $\rho :T M\oplus T^*M \to T M$
\bea
\rho(X+\xi)=\theta (\xi),\label{Sasa anchor}
\ena
and a skew-symmetric bracket
\bea
[X+\xi,Y+\eta ] =[\xi,\eta ]_\theta +\bar{\cal L}_{\xi}Y -\bar{\cal L}_{\eta}X 
-\frac{1}{2}d_\theta (\bar{\iota}_\xi Y  -\bar{\iota}_\eta X ).
\label{Sasa bracket}
\ena
Then the quadruple $(TM\oplus T^*M, \langle \cdot,\cdot \rangle,\rho,[\cdot,\cdot])$ 
defines a Courant algebroid, which
we denote as $(TM)_0\oplus (T^*M)_\theta$ for short.
Its fundamental properties were investigated in \cite{Asakawa:2014kua} 
under the name of a Poisson Generalized Geometry.
There are two kinds of transformations in this algebroid:
$\beta$-transformations 
\bea
	e^\beta(X+\xi)=X+\xi +i_\xi\beta,\label{beta trf}
\ena
where $\beta$ is a bi-vector $\beta \in\Gamma(\wedge^2 T M)$,
and the $\beta$-diffeomorphisms
\bea
	e^{\bar\cL_\zeta } (X+\xi)=e^{\bar\cL_\zeta } X+e^{\bar\cL_\zeta }\xi,\label{beta diff}
\ena
which is generated by any $1$-form $\zeta\in\Gamma(T^*M)$.
A $\beta$-transformation implies the relation
\bea
[e^\beta (X+\xi),e^\beta (Y+\eta) ]
=e^\beta \left( [X+\xi,Y+\eta ] +\bar\iota_\eta \bar\iota_\xi d_\theta \beta\right). \label{beta formula}
\ena
Thus the $\beta$-transformation is a symmetry of the bracket if $\beta$ is $d_\theta$-closed.
In particular, we call a $\beta$-transformation as a $\beta$-gauge transformation if $\beta$ is 
$d_\theta$-exact.
For more details, see \cite{Asakawa:2014kua}%
\footnote{
The condition $\bar\cL_\zeta \theta=0$ for $\zeta$ is needed to preserve the anchor map, 
but it is not required in the following argument.}.

It is shown that the bracket (\ref{Sasa bracket}) satisfies the following relations
\bea
	&[u,fv]
	 	=f [u,v ] +(\bar{\cal L}_{\xi}f) v
		  -(d_\theta f) \langle u,v \rangle,  \label{bracketproperty1}\\
	 &\bar\cL_\xi \langle v , w  \rangle 
	 	= \langle [u,v]+d_\theta  \langle u , v  \rangle  , w  \rangle  
		+  \langle v , [u,w]  +d_\theta  \langle u , w  \rangle\rangle , \label{bracketproperty2}
\ena
where $u=X+\xi$, $v=Y+\eta$, $w=Z+\zeta \in (TM)_0\oplus (T^*M)_\theta$, 
and $f$ is any smooth function%
\footnote{Note that in \eqref{bracketproperty2}
the right-hand side seems to depend on both vector field $X$ and $1$-form $\xi$,
although the left-hand side depends only on $1$-form $\xi$.
We can show, however, that the terms involving $X$ 
in the right-hand side cancel and thus the right-hand side is independent of $X$.}.
They stem from (combinations of) the axioms of a Courant algebroid \cite{Liu97manintriples,Roytenberg}. 
For their proof, 
see appendix \ref{bracketproperty}.
These properties are crucial in defining a connection which is compatible with
$O(n,n)$-invariant inner product $\langle \cdot , \cdot \rangle$ 
as we shall see in the following subsection.

\paragraph{$R$-flux}
In \cite{Asakawa:2014kua}, an $R$-flux is defined as a dual analogue of an $H$-flux 
\cite{Severa:2001qm,Roytenberg:2001am}
in $TM\oplus T^*M$.
It is a trivector $R \in \Gamma(\wedge^3 TM)$ and is an Abelian field strength associated 
with the local $\beta$-gauge symmetry. 
The $R$-flux is written locally by a set of gauge field bivectors 
$\beta_\alpha \in \Gamma(\wedge^2 TU_\alpha)$ as
\bea
R|_{U_\alpha}=d_\theta \beta_\alpha, \label{R-flux def}
\ena
on an open set $U_\alpha$ of $M$.
Note that a $\beta$-gauge transformation $\beta_\alpha \to \beta_\alpha + d_\theta \Lambda_\alpha$
keeps $R$ invariant.

An $R$-flux\footnote{More precisely, its class is in the third Poisson cohomology $[R]\in H_\theta ^3 (M)$.}
is used to deform the structure of the Courant algebroid $(TM)_0\oplus (T^*M)_\theta$, which is called 
an $R$-twisting.
On one hand, $R$ defines an $R$-twisted Courant algebroid $E$ which satisfies the exact sequence  
\bea
0\to (TM)_0 \to E \to (T^*M)_\theta \to 0 ,\label{exact sequence}
\ena
with the same bracket as (\ref{Sasa bracket}).
On the other hand, $R$ defines a Courant algebroid 
$(TM\oplus T^*M, \langle \cdot,\cdot \rangle,\rho,[\cdot,\cdot]_R)$
with the $R$-twisted bracket
\bea
[X+\xi,Y+\eta ]_R=[X+\xi,Y+\eta ] +\bar\iota_\eta \bar\iota_\xi R.\label{twisted Sasa}
\ena
It is shown in \cite{Asakawa:2014kua} that these two deformations are equivalent,
that is, there is an isomorphism $\varphi :TM\oplus T^*M\to E$ of Courant algebroids, 
which comes essentially from the local version of the formula \eqref{beta formula}.
In the following, we adopt the latter description of the $R$-twisting.
When an $R$-flux is present, 
we still denote the Courant algebroid as $(TM)_0\oplus (T^*M)_\theta$ but 
with the $R$-twisted bracket (\ref{twisted Sasa}).

Since the definition and the appearance of an $R$-flux are analogous to those of an 
$H$-flux in standard GG,
we expect that an $R$-flux plays a role analogous to the one of an $H$-flux in the gravity theory. 

\subsection{Generalized Riemannian structure}

As in the standard Generalized Geometry, a generalized Riemannian structure is defined as 
a maximally positive-definite subbundle $C_+$ 
of the bundle $T M  \oplus T^*M$, where the positivity is defined with 
respect to the inner product (\ref{Sasa inner}).
Together with its orthogonal complement $C_-$, we thus have $T M  \oplus T^*M=C_+ \oplus C_-$.

In general, the subbundles $C_\pm $ are given by graphs associated with maps 
$\pm G+\beta : T^*M \to T M$
\bea
C_\pm =\{ \xi +(\pm G+\beta) (\xi)~|~\xi\in T^*M\},
\label{Riem str}
\ena
respectively \cite{Asakawa:2014kua}.
Here the symmetric part $G$ of the maps is a Riemannian metric on $T^*M$, since 
the inner product of elements in $C_+$ reduces to the Riemannian metric on $T^*M$, as
$\langle \xi +(G+\beta) (\xi) , \eta +(G+\beta) (\eta)  \rangle =G(\xi,\eta)$.
The skew-symmetric part $\beta \in \Gamma(\wedge^2 TM)$ is a bivector on $M$.
We can simplify an argument concerning this bivector $\beta$,
by recognizing \eqref{Riem str} as the result of performing a
$\beta$-transformation $e^{\beta}$ (\ref{beta trf}) on the bundle
\bea
C'_\pm =\{ \xi \pm G (\xi)~|~\xi\in T^*M\}.\label{simpler Riem str}
\ena
According to the formula \eqref{beta formula}, there is an isomorphism between these two Courant algebroids,
$e^\beta: C'_+\oplus C'_- \to C_+\oplus C_-$, where the bracket of the former is 
$[X+\xi,Y+\eta ]_{R+d_\theta \beta}$.
Thus, by redefining 
the $\beta$-gauge potential as $\beta_\alpha \to \beta_\alpha +\beta$ 
on each open set $U_\alpha$, 
the effect the bivector $\beta$ is absorbed into a replacement of the $R$-twisted bracket.
In the same manner, 
a further $\beta$-transformation $e^{\tilde{\beta}}$ with $d_\theta \tilde{\beta}=0$ might also be absorbed 
into a redefinition of the bracket, but it does not change 
the bracket at all. 
This is understood as a gauge transformation of the $\beta$-gauge potential.

In this way, a choice of a Riemannian metric $G$ on $T^*M$ is 
translated into a choice of the subbundle $C'_+$
of the form (\ref{simpler Riem str}), which we denote $C_+$ hereafter, omitting the prime.
Note that this observation, 
together with the replacement of the bracket with the twisted bracket, 
plays an important role in the construction of the gravity theory
on the standard Courant algebroid $T M  \oplus T^*M$ \cite{Hitchin:2010qz}.

\paragraph{Projections and lifts}
In order to represent the sections of the subbundles $C_\pm$,
we introduce two kinds of maps \cite{Hitchin:2010qz}, which we call projections and lifts.
We define the lifts $^\pm: \Gamma(T^* M) \to \Gamma(C_\pm)$ 
and $^\pm: \Gamma(TM) \to \Gamma(C_\pm)$, respectively, by
\bea
\xi^\pm:=\xi \pm G(\xi), ~~X^\pm :=X\pm G^{-1}(X),\label{lifts}
\ena
for any $1$-form $\xi$ and vector field $X$.
In the latter, we used the fact that $C_\pm$ are also written as 
$C_\pm =\{ X \pm G^{-1} (X)~|~X\in TM\}$.
The projections $\pi_\pm:\Gamma(T M\oplus T^* M)\to \Gamma(C_\pm)$ are canonically defined.
By noting
$\pi_\pm(X)=\frac{1}{2}\pi_\pm (X+G^{-1}(X)+X-G^{-1}(X))= \frac{1}{2}(X\pm G^{-1}(X))$,
and $\pi_\pm(\xi)=\frac{1}{2}\pi_\pm (\xi+G(\xi)+\xi-G(\xi))=\frac{1}{2}(\xi\pm G(\xi))$,
the projections are written in terms of the lifts as well: 
\bea
	&\pi_\pm (X+\xi ) = \frac{1}{2} (X^\pm + \xi^\pm).\label{projections}
\ena

\subsection{Contravariant connection, torsion and curvature}

In this subsection 
we define a contravariant connection and its torsion and curvature on the
positive-definite subbundle $C_+$ of
the Courant algebroid $(TM)_0\oplus (T^*M)_\theta$ equipped with an $R$-twisted bracket.
The definitions are the contravariant analogue of those given by Hitchin 
for the standard Courant algebroid $T M \oplus T^*M$ \cite{Hitchin:2010qz}.

\paragraph{Contravariant connection}

Let $\bar{D}$ be a bilinear map $\bar{D}: \Gamma(T^*M  \otimes C_+) \to  \Gamma(C_+)$ defined by
\bea
	\bar{D}_\xi (u) := \pi_+ \big( [\xi^-,u]_{R} \big),	\label{Pconnection}
\ena
where $\xi\in \Gamma(T^*M)$ and $u \in \Gamma(C_+)$.
Then, the following properties are satisfied
\bea
	&\bar{D}_{f\xi}(u)=f\bar{D}_\xi (u), \label{connection1} \\
	&\bar{D}_\xi (fu) = f\bar{D}_\xi (u) + (\bar\cL_\xi f)u, \label{connection2}
\ena
for any smooth function $f$ (For a proof, see appendix \ref{connection}.).
Hence, the map $\bar{D}$ 
defines a contravariant connection (\ref{condition of contra connecttion}) 
on the bundle $C_+$.

Furthermore, we can show that
$\bar{D}$ is compatible with the canonical $O(d,d)$-invariant inner product (\ref{Sasa inner}):
\bea
	\bar\cL_\xi \langle u,v \rangle
	=  \langle \bar{D}_\xi (u),v \rangle+ \langle u, \bar{D}_\xi (v) \rangle,	\label{compatible}
\ena
for any $u$ and $v \in \Gamma(C_+)$ (see appendix \ref{compatibility}).

\paragraph{Contravariant torsion}

The torsion associated with the 
contravariant connection $\bar{D}$
can also be defined in a parallel manner with \cite{Hitchin:2010qz}, 
although $\bar{D}$ is not an affine connection.
For $\xi$ and $\eta\in\Gamma(T^*M)$, 
we define 
the contravariant torsion  $\bar \tau : \Gamma(\wedge^2 T^*M )\to \Gamma(C_+)$ by
\bea
	\bar{\tau}(\xi,\eta)=\bar{D}_{\xi} (\eta^+) - \bar{D}_{\eta} (\xi^+) - ([\xi,\eta]_\theta)^+,  \label{319}
\ena
which is actually tensorial since, with the use of the formula \eqref{a20}, it satisfies
\bea
	\bar{\tau}(f\xi,g\eta)=fg \bar{\tau}(\xi,\eta).
\ena

\paragraph{Contravariant curvature}

We define the curvature 
of $\bar{D}$ as a map $\bar{\Omega} : \Gamma(\wedge^2 T^*M )\to \Gamma({\rm End}(C_+))$ by
\bea
	\bar{\Omega}(\xi,\eta)u=
	(\bar{D}_\xi \bar{D}_\eta - \bar{D}_\eta \bar{D}_\xi - \bar{D}_{[\xi,\eta]_\theta})u,\label{bar Omega}
\ena
for $\xi,\eta\in\Gamma(T^*M)$ and $u \in \Gamma(C_+)$.
It satisfies the tensorial property (see appendix \ref{CT}),
\bea
	\bar{\Omega}(f\xi,g\eta)(hu)=fgh\bar{\Omega}(\xi,\eta)u.
	\label{CurvatureTensor}
\ena

In summary, the contravariant connection as well as torsion and the curvature
are consistently defined on $C_+$.
Note that in the proofs of their well-definedness, we merely utilized the properties \eqref{bracketproperty1} 
and \eqref{bracketproperty2}, coming from
the axioms of Courant algebroids.
Thus, these
constructions are also applicable to any Courant algebroid.
To 
elaborate on the characteristic properties
in our case, 
we next examine
 the local expressions of these objects with the full-use of the $R$-twisted bracket.

\subsection{Local expressions}

In this subsection, we present the local expressions of the objects in the previous section.
In particular, we show that the contravariant connection 
$\bar{D}$ on $C_+$ is the lift of a contravariant metric affine connection on $T^*M$.
In local coordinates, we take the coordinate basis as $\{dx^i\}$ for $T^*M$.
Then, by applying the lift, $\{(dx^i)^+=dx^i +G^{ij} \p_j \}$ are the basis of $C_+$.

\paragraph{Contravariant connection}

First we shall calculate the connection \eqref{Pconnection} 
for $\xi=dx^i$ and for $u=(dx^j)^+$,
\bea
	\bar{D}_{dx^i} (dx^j)^+
	&= \pi_+ ( [(dx^i)^-,(dx^j)^+]+\bar{\iota}_{dx^j}\bar{\iota}_{dx^i}R ).\label{local form of connection}
\ena
We can read off the connection coefficients from $\bar{D}_{dx^i} (dx^j)^+=\bar{\Upsilon }^{ij}_k (dx^k)^+$ as
\bea
	\bar{\Upsilon }^{ij}_k
		=\bar{\Gamma}^{ij}_k
		+\frac{1}{2} {R}^{ijn}G_{nk},
	\label{upsilon}
\ena
where $\bar{\Gamma}^{ij}_k$ is the contravariant Christoffel symbol \eqref{220}
of the contravariant Levi-Civita connection,
 and ${R}^{ijn}$ is the $R$-flux.

To show this, we note that the bracket in
the first term in \eqref{local form of connection}
is evaluated by using (\ref{Sasa bracket}) as\footnote{
For computational details, see appendix \ref{sbracketbeta0}.}
\bea
	&[dx^i-G^{ik}\p_k , dx^j+G^{jl}\p_l]\nn
	&= \p_k \theta^{ij}dx^k 
		+ [\theta^{mn}(\p_m G^{ji})  
		-\theta^{mi}(\p_m G^{jn})  -\theta^{mj}(\p_m G^{in})
		 -  G^{jl} ( \p_l \theta^{in})  
		 -  G^{il} ( \p_l \theta^{jn})] \p_n. \label{bracketbeta0}
\ena
Acting the projection $\pi_+$ on the above expression and noting (\ref{projections}), which implies
$\pi_+ (dx^k) = \frac{1}{2} (dx^k)^+$ and $\pi_+ (G^{nk} \p_n)= \frac{1}{2}(dx^k)^+$,
the first term of the connection coefficients \eqref{local form of connection} is nothing but
the contravariant Christoffel symbol \eqref{220}:
\bea
	\pi_+ &( [(dx^i)^-,(dx^j)^+] \nn
     =&\frac{1}{2} [\p_k \theta^{ij}+
	 G_{nk}(\theta^{mk}\p_m G^{ij}-\theta^{mi}\p_m G^{jk}-\theta^{mj}\p_m G^{ki}
	  -G^{jm}\p_m\theta^{ik} -G^{im}\p_m\theta^{jk})] (dx^k)^+ \nn
	  =&\bar{\Gamma}^{ij}_k (dx^k)^+.
\ena
Note that the metric compatibility (\ref{28}) of $\bar \nabla$ is equivalent to 
\eqref{compatible}.
Thus, when the $R$-flux is absent, the contravariant connection $\bar{D}$ on $C_+$ 
is the lift of the contravariant Levi-Civita connection $\bar \nabla$ on $T^*M$.

Since $R$ in (\ref{local form of connection}) is a trivector%
\footnote{Because $R$ is written by a local bivector potential as $R=d_\theta \beta_\alpha =[\theta,\beta_\alpha]_S$, 
its components are also written as 
\bea
	R^{ijn}
	  &=
	  \theta^{nm}\p_m\beta_\alpha^{ij}
	  +\theta^{im}\p_m\beta_\alpha^{jn}
	  +\theta^{jm}\p_m\beta_\alpha^{ni}
	  +\beta_\alpha^{nm}\p_m \theta^{ij}
	  +\beta_\alpha^{im}\p_m\theta^{jn}
	  +\beta_\alpha^{jm}\p_m\theta^{ni}.\nonumber
\ena},
$R=\frac{1}{3!}R^{ijn}\p_i \wedge \p_j \wedge \p_n $, we can straightforwardly evaluate 
the second term of (\ref{local form of connection}) as
\bea
\pi_+ (\bar{\iota}_{dx^j}\bar{\iota}_{dx^i}R) 
&=\pi_+ (R^{ijn}\p_n)
=R^{ijn}G_{nk}\pi_+ (G^{kl}\p_l ) 
=\frac{1}{2} R^{ijn}G_{nk} (dx^k)^+.
\ena

In summary, we obtain the explicit form of the contravariant connection on $C_+$ as
\bea
	\bar{D}_{dx^i} (dx^j)^+
		=\bigg(\bar{\Gamma}^{ij}_k
		+\frac{1}{2} {R}^{ijn}G_{nk}\bigg)(dx^k)^+.
	\label{conexpform11}
\ena
It is the lift of a contravariant metric affine connection on $T^*M$.

\paragraph{Contravariant torsion}

The components of the torsion tensor (\ref{319})
is obtained by computing
\bea
	\bar{\tau}(dx^i,dx^j)=\bar{\tau}_k^{ij}(dx^k)^+,
\ena
and is written in terms of $\bar{\Upsilon }^{ij}_k$ in \eqref{upsilon} as
\bea
	\bar{\tau}_k^{ij}= \bar{\Upsilon }^{ij}_k-\bar{\Upsilon }^{ji}_k - \p_k\theta^{ij}.
\ena
This implies that the torsion tensor of $\bar{D}$ is also the lift of
the torsion tensor $\bar{T}$ (\ref{2.15}) of the contravariant affine connection $\bar{\nabla}$.
In fact, from (\ref{upsilon}) we have
\bea
	\bar{\tau}_k^{ij}
    &=(\bar{\Gamma}^{ij}_k-\bar{\Gamma}^{ji}_k
	- \p_k\theta^{ij})  +{R}^{ijn}G_{nk} \nn
&={R}^{ijn}G_{nk},
\ena
where (\ref{Ttheta}) is used.
Thus, the contravariant Levi-Civita part of the connection $\bar{\Upsilon }^{ij}_k$ 
does not contribute to the torsion $\bar \tau$, 
whereas the $R$-flux does.
This is parallel to the case of the standard generalized geometry, where
an $H$-flux appears as the torsion of the generalized Bismut connection \cite{GualtieriPoisson}.

\paragraph{Contravariant curvature}

The components of the curvature tensor \eqref{bar Omega} are found by 
\bea
	\bar{\Omega}({dx^i},{dx^j})(dx^k)^+ =\bar{\Omega}^{kij}_l (dx^l)^+,
\ena
and are written in terms of (\ref{upsilon}) as
\bea
	\bar{\Omega }^{kij}_{l}
	&= \theta^{im} \p_m \bar{\Upsilon }^{jk}_l 
		- \theta^{jm} \p_m \bar{\Upsilon }^{ik}_l -\p_n \theta^{ij} \bar{\Upsilon }^{nk}_l 
		+ \bar{\Upsilon }^{jk}_m \bar{\Upsilon }^{im}_l 
		 - \bar{\Upsilon }^{ik}_m \bar{\Upsilon }^{jm}_l .\label{bar Omega local}
\ena
Of course, for the case of a vanishing $R$-flux ($R^{ijn}=0$ in \eqref{upsilon}), 
this reduces to the contravariant Riemann tensor \eqref{contra Riem}:
\bea
	\bar{R}^{kij}_{l}
	&= \theta^{im} \p_m \bar{\Gamma}^{jk}_l 
		- \theta^{jm} \p_m \bar{\Gamma}^{ik}_l -\p_n \theta^{ij} \bar{\Gamma}^{nk}_l 
		+ \bar{\Gamma}^{jk}_m \bar{\Gamma}^{im}_l 
		 - \bar{\Gamma}^{ik}_m \bar{\Gamma}^{jm}_l .
\ena
The remaining terms in \eqref{bar Omega local} should be written by a combination of tensors.
Explicitly, we find\footnote{See, Appendix \ref{compdet}.}
\bea
	\bar{\Omega}^{kij}_l 
	=& \bar{R}^{kij}_{l}
+ \frac{1}{2}\left( \bar \nabla_{dx^i} {R}^{jkn} - \bar \nabla_{dx^j} {R}^{ikn}\right) G_{nl} 
+ \frac{1}{4}G_{nm}\left( {R}^{jkn}{R}^{imp} -{R}^{ikn}{R}^{jmp}\right)G_{pl}.\label{gene Riem}
\ena

\paragraph{Generalized contravariant Ricci tensor and Ricci scalar}

Since the index structure of $\bar{\Omega }^{kij}_{l}$ is the same as the contravariant Riemann tensor,
we define the corresponding Ricci tensor and the Ricci scalar 
in the same way as \eqref{contra Ricci} and 
\eqref{contra Ricci scalar}, respectively, prefixing the term ``generalized'' with them:
The generalized contravariant Ricci tensor $\bar{\Omega}^{kj}$ is obtained from \eqref{gene Riem} as
\bea
	\bar{\Omega}^{kj}:=&\bar{\Omega}^{klj}_l\nn
	=& \bar{R}^{kj} 
+ \frac{1}{2}\bar \nabla_{dx^l} {R}^{jkn}G_{nl} 
+ \frac{1}{4}G_{nm}\left( {R}^{jkn}{R}^{lmp} -{R}^{lkn}{R}^{jmp}\right)G_{pl} ,
\ena
with $\bar{R}^{kj}$ being \eqref{contra Ricci};
the generalized contravariant Ricci scalar $\bar{\Omega}$ is given by
\bea
	\bar{\Omega}:=& G_{kj}\bar{\Omega}^{kj}	=
	\bar{R}  -\frac{1}{4}{R}^2, \label{PoigeneralizedRicci}
\ena
where $\bar R$ is given in \eqref{contra Ricci scalar} and $R^2$ denotes
\bea
	&{R}^2=G_{il}G_{jm}G_{kn}{R}^{ijk}{R}^{lmn}.
\ena

Thus the generalized contravariant
Ricci scalar $\bar{\Omega}$ is given by a sum of the contravariant Ricci scalar $\bar{R}$,
and the square of the $R$-flux.
Since the contravariant Levi-Civita part is defined independently of the presence of $R$-fluxes, 
we may consider a particular case of $R=0$ without any problem.
The outcome \eqref{PoigeneralizedRicci}
 has the same structure as the generalized Ricci scalar 
obtained in standard Generalized Geometry,
where the generalized Ricci scalar consists of the ordinary 
Ricci scalar and the square of the $H$-flux, see appendix \ref{RGGG}.

\subsection{Invariant measure}

Towards the construction of an action for a gravity theory that is relevant to our geometry,
we would like to determine an appropriate invariant measure,
or equivalently a volume form. 
In this subsection, we argue that if the Poisson manifold $M$ is unimodular, 
there is a measure such that the integrations of scalar functions 
are invariant under $\beta$-diffeomorphisms.

To clarify this, we shall recall first 
the notion of an invariant measure in the ordinary general relativity.
If we assume that the $n$-dimensional manifold $M$ is orientable, 
we can always choose a volume form (nowhere vanishing top form) 
$\Sigma   \in \Gamma(\wedge^n T^*M)$, written in local coordinates,
\bea
\Sigma  =\rho dx^1 \wedge \cdots \wedge dx^n.\label{volume form}
\ena
Under a diffeomorphism generated by a vector field $X=X^i\p_i$, it transforms as
\bea
	\cL_X \Sigma  = d( i_X \Sigma) = \p_i (\rho X^i ) dx^1\wedge \cdots \wedge dx^n.\label{diff sigma}
\ena
which is a total divergence
 ($d$-exact top form).
Hence, the integration of a scalar function $f$ (i.e., $\rho f$ is a Lagrangian density) 
over the manifold $M$ is diffeomorphism invariant
\bea
\int_M {\cal L}_X (\Sigma f)= \int  \p_i (\rho X^i f) d^n x =0,
\ena
provided that the surface integral vanishes.
In particular, if $X$ is divergence free, $div_\Sigma X=\frac{1}{\rho} \p_i (\rho X^i  )=0$, 
the volume form $\Sigma$ itself is invariant.
On a Riemannian manifold $M$ equipped with a metric $G$, 
it is conventional to set 
\bea
	\rho = \sqrt{\det G} .
\ena

In our case of the gravity on a Poisson manifold,
we would like to introduce a volume form $\Sigma $ 
such that its $\beta$-diffeomorphism $\bar{{\cal L}}_\zeta \Sigma$ 
reduces to a total divergence.
However, it is in general impossible to find such $\Sigma $, 
because $\bar{{\cal L}}_\zeta \Sigma  $ is not necessarily $d$-exact. 
Explicitly, we have
\bea
\bar{{\cal L}}_\zeta \Sigma 
&= -\frac{1}{\rho}\p_i (\rho \theta^{ij}) \zeta_j \Sigma  + \theta^{ij} \p_i \zeta_j \Sigma ,
\label{beta on volume}
\ena
and we immediately recognize that the relative sign prevents the right-hand side from being 
a total divergence.

It is known that the modular class of a Poisson manifold $M$ is an obstruction to the existence of a 
volume form on $M$, which is invariant under the action of Hamiltonian vector fields \cite{Weinsteinmodular},
 see also \cite{KSmodular} and references therein.
Since a Hamiltonian vector field $X_f =-d_\theta f $ is a particular case of the $\beta$-diffeomorphism,
it is natural to expect that there is a desired volume form when the modular class is trivial.
We show that this is indeed the case.

\paragraph{Modular class}

Given a volume form $\Sigma $ as (\ref{volume form}), 
by applying the formula (\ref{diff sigma})  to a Hamiltonian vector field 
$X_f=-d_\theta f =\theta(df)$, we have
\bea
{\cal L}_{X_f}\Sigma  =X_\Sigma  (f) \Sigma ,~~
X_\Sigma  (f)=-\frac{1}{\rho}\p_i (\rho \theta^{ij})\p_j f.
\ena
Since $f$ is arbitrary, 
it is shown that $X_\Sigma $ defines a vector field, 
called the modular vector field, locally represented as
\bea
X_\Sigma =-\frac{1}{\rho}\p_i (\rho \theta^{ij})\p_j .
\ena

Let $\Sigma_1$ be another volume form.
Then, since it can be written as $\Sigma_1 =e^\chi \Sigma$ for some function $\chi$, 
the corresponding modular vector field $X_{\Sigma_1} $ is given by 
\bea
X_{\Sigma_1} = X_{\Sigma} + d_\theta \chi.
\label{modular diff}
\ena
It is also shown that $d_\theta X_\Sigma =0$.
Therefore, its cohomology class $[X_\Sigma] \in H_\theta^1 (M)$, called the modular class, 
in the first Poisson cohomology is independent of the volume form \cite{Weinsteinmodular}.\\
{\it Proof.}
The left hand side of the equation ${\cal L}_{X_f}\Sigma_1  =X_{\Sigma_1}  (f) \Sigma_1$ becomes
\bea
{\cal L}_{X_f} \Sigma_1
&={\cal L}_{X_f} (e^\chi \Sigma)
=({\cal L}_{X_f} \chi ) e^\chi \Sigma +e^\chi ({\cal L}_{X_f} \Sigma )\nn
&=\theta (df,d\chi ) \Sigma_1 + X_{\Sigma} (f) \Sigma_1
=\left( X_{\Sigma} (f) -\theta (d\chi, df) \right)\Sigma_1 \nn
&=\left( X_{\Sigma} +d_\theta \chi  \right) (f) \Sigma_1.
\ena
This shows the first assertion \eqref{modular diff}.
Next, the equation
\bea
X_{\Sigma} (\{ f,g\} ) \Sigma
&={\cal L}_{X_{\{ f,g \}} }\Sigma 
=[{\cal L}_{X_f}, {\cal L}_{X_g}]\Sigma \nn
&={\cal L}_{X_f} X_{\Sigma} (g) \Sigma - {\cal L}_{X_g} X_{\Sigma} (f) \Sigma  \nn
&=\left( X_f (X_{\Sigma} (g)) - X_g (X_{\Sigma} (f)) \right) \Sigma \nn
&=\left( \{ f, X_{\Sigma} (g)\} + \{ X_{\Sigma} (f),g\} \right) \Sigma,
\ena
implies that the vector field $X_{\Sigma}$ generates an infinitesimal flow preserving the Poisson bracket.
It means that $0={\cal L}_{ X_{\Sigma}} \theta =d_\theta X_{\Sigma}$.
{\it q.e.d.}

Note that the equation (\ref{beta on volume}) is rewritten by using the modular vector field as
\bea
\bar{{\cal L}}_\zeta \Sigma 
&=  \left(i_\zeta X_\Sigma + \theta^{ij} \p_i \zeta_j \right) \Sigma .
\label{beta on volume2}
\ena

\paragraph{Unimodular case}

In general, a Poisson manifold $M$ is called unimodular%
\footnote{It is different from the unimodular condition $\sqrt{-g}=1$ used in the unimodular gravity.},
if its modular class vanishes, $[X_\Sigma]=0$.
In this case, a modular vector field is always $d_\theta$-exact, that is, 
$X_\Sigma =d_\theta \phi$ for some function $\phi$ on $M$:
\bea
	  X_\Sigma= -\frac{1}{\rho}  \p_i(\rho \theta^{ij}   ) \p_j= -\theta^{ij}\p_i\phi \p_j. \label{unimo}
\ena

Let $\Sigma_1 =e^{-\phi} \Sigma$ be another volume form induced from $\Sigma$.
Then, according to (\ref{modular diff}), the corresponding modular vector field vanishes: $X_{\Sigma_1}=X_\Sigma -d_\theta \phi =0$.
Hence, $\Sigma_1$ is invariant  ${\cal L}_{X_f}\Sigma_1 =0$ for any Hamilton vector field $X_f$.

Moreover, we observe that the same volume form $\Sigma_1$ transforms as a total divergence
under a $\beta$-diffeomorphism.
Indeed, by applying (\ref{beta on volume2}) for $\Sigma_1$, we have 
\bea
\bar{{\cal L}}_\zeta (\Sigma_1 )
&=\theta^{ij} \p_i \zeta_j \Sigma_1 \nn
&=\p_i ( \rho \theta^{ij} e^{-\phi} \zeta_j) dx^1\wedge \cdots \wedge dx^n,
\ena
where we used 
\bea
\p_i ( \rho \theta^{ij} e^{-\phi} \zeta_j) 
&=\p_i (\rho \theta^{ij} ) e^{-\phi} \zeta_j -\rho \theta^{ij} \p_i \phi e^{-\phi} \zeta_j 
+\rho \theta^{ij} e^{-\phi} \p_i \zeta_j \nn
&=\rho \theta^{ij} e^{-\phi} \p_i \zeta_j,
\ena
following from (\ref{unimo}). Hence, the integral defined by the
measure ${\Sigma}_1$ is invariant under $\beta$-diffeomorphisms.

Since in our case, a unimodular Poisson manifold $M$ 
is also a Riemannian manifold equipped with a metric $G$, 
we choose the Riemannian volume form $\rho=\sqrt{\det G}$ in $\Sigma$.
Then, our invariant measure is
\bea
	\Sigma_1 = e^{-\phi}\sqrt{\det G} dx^1 \wedge \cdots \wedge dx^n.\label{inv mesure}
\ena
The integration of any scalar function $f$ with this measure
is invariant under both diffeomorphism and $\beta$-diffeomorphism.
This factor $e^{-\phi}$ is a reminiscent of the dilaton field.
However, it should be determined by the partial differential equation 
\bea
	 \theta^{ij}\p_i\phi =\frac{1}{ \sqrt{\det G}}  \p_i (\sqrt{\det G} \theta^{ij}   ) ,\label{358}
\ena
which follows from (\ref{unimo}). The existence of the solution is guaranteed by the unimodularity.

As a special case, we consider a symplectic manifold, which is a unimodular Poisson manifold.
In this case, (\ref{358}) reduces to
\bea
	\p_k\phi& 
	=-G_{ij}\p_k G^{ij} + \theta^{-1}_{jk} \p_i \theta^{ij}. 
\ena
For the case of a K\"ahler manifold in particular, it would be an interesting question 
how this scalar field $\phi$ controls the balance of the metric and the symplectic form.
In \cite{Boucetta:2011ofa1,Boucetta:2011ofa2,Boucetta:2011ofa3}, it is also shown that a Riemann-Poisson manifold is unimodular.

\subsection{Einstein-Hilbert-like action and others}

Since we have obtained both the generalized contravariant Ricci scalar $\bar \Omega$
\eqref{PoigeneralizedRicci}
and the invariant measure $\Sigma_1$ \eqref{inv mesure},
an Einstein-Hilbert-like action can be written as
\bea
	S [G,\beta]&= \int d^nx e^{-\phi}\sqrt{\det G_{ij}}
	\bigg(\bar{R}  -\frac{1}{4}{R}^2 \bigg) \label{EHL},
\ena
where $\bar{R}$ is the contravariant Ricci scalar and $R$ is the $R$-flux
obtained in the preceding subsections.
As argued previously, this action is invariant under $\beta$-diffeomorphisms.
Moreover, it is also invariant under $\beta$-gauge transformations.
This is because a $\beta$-gauge transformation can only affect an $R$-flux, 
but it keeps $R$ invariant, as mentioned below \eqref{R-flux def}.
Thus we have obtained a kind of gravity theory that consistently couples with an $R$-flux.
By construction, the dynamical variables of this theory are the metric $G$ and the local bivectors
$\beta_\alpha$.

As in the case of ordinary gravity, this choice of action is just the simplest one.
More generally, 
any scalar (other than $\bar \Omega$) is allowed to be utilized in the construction of an action. 
Since there is the Poisson tensor $\theta^{ij}$ in addition to the metric $G_{ij}$,
the number of possibilities to make scalar quantities is much greater than 
that in the usual gravity theory.
Furthermore, if the inverse of the Poisson tensor exists, 
the degrees of freedom of making a scalar increase tremendously.
For example,
multiplying $\theta^{-1}$ twice on the contravariant Riemann tensor $\bar{R}^{kij}_{l}$ 
\eqref{riempois} yields the ordinary Riemannian tensor ${\sf R}^k_{lab}$ (plus
 other terms made of the contravariant contorsion):
\bea
  &-\theta^{-1}_{ai}\theta^{-1}_{bj}\bar{R}^{kij}_{l} \nn
=& {\sf R}^k_{lab}
		+ (\nabla_n \theta^{-1}_{ab}) K^{nk}_l
 		+\theta^{-1}_{ai} \nabla_b K^{ik}_l 
		 -\theta^{-1}_{bj} \nabla_a K^{jk}_l
		-\theta^{-1}_{ai}\theta^{-1}_{bj}K^{jk}_m  K^{im}_l 
		+\theta^{-1}_{ai}\theta^{-1}_{bj}  K^{ik}_mK^{jm}_l  .
\ena
This tensor can also be used to construct the action, 
which might be interpreted as a deformation of the 
ordinary gravity theories including the ordinary Einstein-Hilbert action.
Of course, which action it favors depends on the situation.  
This will be clarified by studying classical solutions of the theory and their geometry, 
or relations to other theories such as string theory.

\section{Conclusion and Discussion}

In this paper, we studied the 
Riemannian geometry on a Poisson manifold,
first in the framework of a 
geometry based on the Lie algebroid
$(T^*M)_\theta$ of a Poisson manifold 
and then in the framework of the Poisson Generalized Geometry based on the Courant algebroid 
$(TM)_0 \oplus (T^*M)_\theta$.
We found the consistent definitions of the contravariant versions of the
connection, torsion and  the curvature tensors in both frameworks 
and established the relationship between them.
It is worth noting that
the $R$-flux appeared as a contravariant torsion tensor.
Then we showed that gravity theories coupled with $R$-fluxes 
can be constructed. 
In particular, using the 
contravariant Ricci scalar $\bar\Omega$,
which is a sum of the contravariant Ricci scalar $\bar R$ and the square of the R-flux as
\bea
	\bar{\Omega}=\bar{R}  -\frac{1}{4}R^2, 
\ena
together with the invariant measure on unimodular Poisson manifolds,
we constructed the Einstein-Hilbert-like action
\bea
	S[G,\beta] = \int d^nx e^{-\phi}\sqrt{\det G}
	\bigg(\bar{R}  -\frac{1}{4}{R}^2 \bigg),
\ena
which is invariant under both $\beta$-gauge transformations and $\beta$-diffeomorphisms.

\paragraph{Comparison with ordinary gravity}
The structure of the results obtained above is the same by construction as those in the case of the standard Generalized Geometry. 
There, the $H$-flux appears as a torsion tensor and the generalized Ricci scalar is a sum of the ordinary Ricci scalar ${\sf R}$
 and the square of the $H$-fluxes:
\bea
	\Omega&
	={\sf R} - \frac{1}{4}H^2,
\ena
where the Ricci scalar ${\sf R}$ is the one constructed from the Riemannian metric $g_{ij}$ and 
 $H^2=g^{il}g^{jm}g^{kn}H_{ijk}H_{lmn}$ (see Appendix \ref{RGGG}). 
The resulting action  
is that of the NSNS sector of the supergravity theory%
\footnote{We ignore the dilaton term.} 
\bea
	S_{\rm NSNS}[g,B] =\int dx^n \sqrt{\det g_{ij}}\bigg({\sf R}-\frac{1}{4}H^2 \bigg),
\label{NSNS}
\ena
which is invariant under both $B$-gauge transformations and diffeomorphisms.

Although these two theories, $S[G,\beta]$ and $S_{\rm NSNS}[g,B]$,
are similar in their structures, they are definitely different in many ways.
The differences 
stem from the difference between two Lie algebroids $(T^*M)_\theta$ and $TM$,
and the two different Courant algebroids $(TM)_0\oplus (T^*M)_\theta$ and $TM\oplus T^*M$.
In particular, $R$-fluxes and $H$-fluxes are generically independent notions.
The difference becomes more apparent if we switch the description of an $R$-flux
via the $R$-twisted bracket to the description via the $R$-twisted Courant algebroid $E$
(see \eqref{exact sequence}). 
An $H$-flux also has such a description by an $H$-twisted Courant algebroid $E'$.
These two algebroids
 $E$ and $E'$ are different as vector bundles, since $E$ is glued by local $\beta$-gauge transformations,
while $E'$ is glued by local $B$-gauge transformations.
Thus there is in general no way to relate 
both of them, except for imposing a particular constraint on the geometry on $M$.

Nevertheless, it is meaningful to compare the two theories, when we focus on the gravitational part only.
This makes sense because the metric structure is independent of whether the fluxes are present or not,
as explained in the beginning of section 3.2.
In particular, we shall assume that the Courant algebroid is twisted by a trivial element of the Poisson cohomology,
or equivalently, the $R$-twisted bracket is given by $R=d_\theta \beta$ with a global bivector $\beta$.
Then, we can come back to the bundle 
\eqref{Riem str} parametrized by $G+\beta$.
Similarly, if the $H$-flux is trivial in a sense similarly as discussed above, $C_+$ is given by a 
graph associated with a map 
$g+B:T M \to T^*M$,
\bea
C_+ =\{ X +(g+B) (X)~|~X\in T M\},
\label{metric from T M}
\ena
where $B$ is a global $2$-form.

Now we recall the argument 
in \cite{Asakawa:2014kua}.
Since the definition of the subbundle $C_+$
depends only on the bundle and the bilinear form, 
a generalized Riemannian structure of the Courant algebroid $(T M)_0  \oplus (T^*M)_\theta$
is equivalent to that
of the standard Courant algebroid $T M  \oplus T^*M$.
In other words, these two Courant algebroids share the same generalized Riemannian structure $C_+$.
The two representations (\ref{Riem str}) and (\ref{metric from T M}) of $C_+$ are related by 
\bea
G+\beta =(g+B)^{-1}.
\ena

Thus, it is possible to compare the two theories 
$S[G,\beta]$ and $S_{\rm NSNS}[g,B]$ through this relation.
Note, however, that the Poisson tensor $\theta$ does not appear in this relation%
\footnote{Note that there is also another relation of two Riemannian metrics $G$ and $g$ 
including the Poisson tensor $\theta$ \cite{Asakawa:2012px,Asakawa:2014eva}, 
appears in a different situation, where both descriptions are 
in the the same Courant algebroid $T M  \oplus T^*M$.}.
One possibility is to treat the Poisson tensor as a matter field as in the argument in \S\ref{cova-contra}.

\paragraph{Future directions}

To understand the dynamics of the theory, we need to investigate the equations of motion, 
and to analyze their classical solutions.
Solutions of particular interest
are the analogue of the fundamental string solution and the NS5-brane solution
that should be charged 
under the $R$-flux,
since such objects are speculated to exhibit exotic geometrical properties
\cite{KOS,OS,Sakatani:2014hba,BS,KS,KSY1,KSY2,Kimura:2014wga,CGMZ,Andriot:2014uda,Park:2015gka}.
An open question is whether our framework provides a 
geometrical meaning for those non-geometric objects.

It is also interesting to study the concept of T-duality within this theory, 
together with the Kalza-Klein reduction.
As shown in \cite{Asakawa:2015aia}, at the topological level,
the $Q$-fluxes appear in PGG when the spacetime $M$ is non-trivially fibered by $S^1$.
At the level of the Riemannian geometry, 
this information should be included in the contravariant scalar curvature term in the action.

There is another open question about the status of the Poisson structure.
By construction, in this paper we assume that the Poisson structure is given before 
a Riemannian metric is defined. 
The resulting equation of motion of the gravity theory determines the metric for a fixed Poisson structure, 
rather than being determined simultaneously. 
Note that our setting is close to the concept of the non-commutative geometry.
In general, a Poisson manifold can be regarded 
as the semi-classical approximation of a non-commutative space,
that is, a Poisson tensor is a part of the characterization of a space.
Hence, by applying Kontsevich's deformation quantization 
 \cite{Kontsevich:1997vb,Cattaneo:1999fm},
 it is natural to expect that
the gravity theory 
in this paper might be lifted to a gravity theory on a non-commutative space.
In this sense, it is interesting to compare our theory with the work of \cite{Aschieri:2005yw}, or 
with the emergent gravity approach in matrix models \cite{Yang,Steinacker1,Steinacker2}.

\paragraph{Comparison with other approaches}
We give a few comments on the relationships to other approaches 
to gravity theory with non-geometric fluxes.

The approach \cite{Blumenhagen:2012nt} is conceptually similar to the content of \S 2 of this paper.
The authors consider a quasi-Poisson version of our Lie algebroid $(T^*M)_\theta$ 
and regard their $R$-flux as the violation of the Poisson condition due to the quasi-Poisson structure.
Based on these notions of geometry,  they construct the contravariant Levi-Civita connection and
obtain an action, which is invariant under their $\beta$-diffeomorphisms.
Their action
is equivalent to \eqref{NSNS} by a field redefinition $G+\beta$ to $g+B$, for the symplectic case.
In particular, eq. (4.23) of \cite{Blumenhagen:2012nt} is formally the same as 
ours \eqref{220}, if we identify their $\beta$ as our $\theta$.

The series of papers by Andriot et.al.
\cite{Andriot:2011uh,Andriot:2012wx,Andriot:2012an,Andriot:2012vb,Andriot:2013xca,Andriot:2013txa,Andriot:2014uda,Andriot}
is conceptually very different from ours, because their formalism is 
based on
the double field theory (DFT) approach
(see \cite{Hull:2009sg,Hull:2009zb,Zwiebach:2011rg,Aldazabal:2013sca,Hohm:2013bwa} for DFT.).
They consider differential geometry on the doubled space 
and construct a connection. 
The resulting theory is a sum of the ordinary and the dual Ricci scalar plus non-geometric flux terms,
which is manifestly invariant under the ordinary diffeomorphism.
By dropping the dependence on the dual coordinates, 
their outcome is also equivalent to \eqref{NSNS} \cite{Andriot:2013xca}.
In this approach, there are also similar formulae to ours.
In particular, eq. (24) in \cite{Andriot:2012wx} is 
formally the same as 
our \eqref{220}, 
our relations (2.46) and (2.47)  are given in (1.26) of \cite{Andriot:2013xca}
and it is also suggested that our (2.48) corresponds to (3.13) of \cite{Andriot:2013xca}%
\footnote{These points are clarified in communication with David Andriot.},
if we again identify their $\beta$ as our $\theta$.

However, there are significant differences 
between the work mentioned above and our construction that can be summarized as follows:
In both approaches,\cite{Blumenhagen:2012nt} and \cite{Andriot:2011uh,Andriot:2012wx,Andriot:2012an,Andriot:2012vb,Andriot:2013xca,Andriot:2013txa,Andriot:2014uda}
our Poisson tensor $\theta$ is replaced by a (not necessary Poisson) bivector $\beta$.
This causes a breakdown of the Lie algebroid structure of $(T^*M)_\theta$. 
As a result, they need to introduce a 
rather complicated definition of $\beta$-diffeomorphism and the notion of 
$\beta$-tensors.
We expect that our construction in this paper should have clarified this point 
and achieved an improvement in geometrical understanding.

Furthermore, it should be noted that
their definitions of $Q$- and $R$-fluxes, are different from ours.
In both approaches, a bivector $\beta$ 
plays the role of a source of such fluxes.
For example, they are defined in \cite{Andriot:2012wx} as
\bea
	&R^{ijk}= 3(\tilde{\p}^{[i}\beta^{jk]}+\beta^{l[i}\p_l\beta^{jk]}),~~
	Q^{~ij}_k =\p_k \beta^{ij},
\ena
and by dropping the dual derivative, they are equivalent to those in \cite{Blumenhagen:2012nt}.
$R^{ijk}$ appears in the commutator of the connection, and $Q^{~ij}_k$ appears in the 
skew-symmetric part of the connection\footnote{
In \cite{Andriot:2012wx,Andriot:2014uda}, 
a differnt definition of a $Q$-flux is given, in terms of  flat indices.
}.
This suggests again that their fluxes correspond to the gravitational part of our theory (\S 2 in this paper),
by the replacement of our $\theta$ with their $\beta$.
If $\beta=\theta$ is Poisson, then $R^{ijk}=0 $, which is guaranteed by the Poisson condition $[\theta,\theta]_S=0$,
and the $Q^{~ij}_k$ above is nothing but the skew-symmetric part of our contravariant Levi-Civita connection.
As a result, 
it is not clear whether an $R$-flux in their definition is a gauge field strength like an $H$-flux,
and it is not explained why the combination of the Ricci scalar and the $R$-square term appears.

On the other hand, in our approach, the roles of the Poisson tensor $\theta$ and the bivector 
potential $\beta$ are 
independent of each other, and the $R$-flux is treated in the same way as an $H$-flux.
That is, our $R$-flux is independent of the gravity part (i.e. the contravariant Levi-Civita connection), 
but is combined with the gravity as a torsion when considering a connection on $C_+$. 
This explains the combination above as a generalized Ricci scalar,
although the physical meaning of $\theta$ is still unclear.
In our approach, a $Q$-flux appears also in the different places:
it is associated with the Kaluza-Klein reduction of a metric \cite{Asakawa:2015aia}.

Another but the most important difference lies in the point where
they consider that the
theories should be equivalent to 
the ordinary gravity theory with an $H$-flux, whereas we do not assume such an equivalence.
However, rather than focusing on such 
differences, in our opinion, 
it would be better to pursue the study of their similarities more concretely.
For example, the formal coincidence to the formula (24) in \cite{Andriot:2012wx}, mentioned above, 
suggests that our contravariant Levi-Civita connection can be lifted to the DFT framework.
Then, it is interesting to see how our $R$-flux is lifted under such a scenario.
After lifting to the DFT framework, it would be possible to compare these approaches
with ours in more detail.


\section*{Acknowledgments}
Authors would like to thank 
the members of the particle theory and cosmology group, 
in particular U.~Carow-Watamura for helpful comments and discussions.
They would also like to thank David Andriot for valuable comments
and suggestions.
H.~M. is supported in part by the Tohoku University
Institute for International Advanced Research and Education.

\appendix

\section{Riemannian geometry based on Generalized Geometry \label{RGGG}}

Here we review on the Riemannian geometry in the framework of the 
standard generalized geometry, based on the Courant algebroid $T M \oplus T^*M$.
For more details, see \cite{David2007}.

\subsection{Courant algebroid $T M \oplus T^*M$}

The Courant bracket in the standard generalized geometry is defined by
\bea
	[X+\xi,Y+\eta]=[X,Y]+\cL_X \eta - \cL_Y \xi - \frac{1}{2} d(i_X \eta - i_Y \xi),
\ena	
where $X,Y\in \Gamma(T M)$ and $\xi,\eta \in \Gamma(T^* M)$.
For $u=X+\xi$, $v=Y+\eta$ and a $2$-form $B$, it is shown that
\bea
	[e^B(u),e^B(v)]&=[u+B(X),v+B(Y)] 
	=e^B([u,v])+ i_Yi_XdB.
\ena
Thus the bracket is invariant under the $B$-transformation $e^B$ with a closed $2$-form, $dB=0$.
In general, in the case of $dB\neq0$, it acts as a shift $H\to H+dB$ in the $H$-twisted bracket
\bea
	[u,v]_H
	&=[u,v]+ i_Yi_X H.
\ena
The Courant bracket satisfies following relations,
\bea
	&[u,fv]
	 	=f [u,v ] +({\cal L}_{X}f) v
		  -d f \langle u,v \rangle,  \label{bracketproperty11}\\
	 &\cL_X \langle v , w  \rangle 
	 	= \langle [u,v]+d  \langle u , v  \rangle  , w  \rangle  
		+  \langle v , [u,w]  +d  \langle u , w  \rangle\rangle , \label{bracketproperty12}
\ena
for $u=X+\xi$, $v=Y+\eta$ and $w=Z+\zeta$,
which are the analogues of \eqref{bracketproperty1} and
\eqref{bracketproperty2}.

\subsection{Generalized Riemannian geometry}

Let $C_+$ be a generalized Riemannian structure of $TM\oplus T^*M$, written 
as a graph $C_+=\{X+g(X)|X\in TM\}$, where $g$ is a Riemannian metric.
A generalized connection, which is a connection 
$D: \Gamma(C_+)\to \Gamma(T^*M\otimes C_+)$ on the vector bundle $C_+$, is obtained by setting
\bea
	D_X u :=  \pi_+( [X^-,u]_{H}) ,
\ena
for any $X\in \Gamma(T M)$ and $ u=Y^+ \in \Gamma(C_+)$.
Here we introduced lifts $^\pm$ as $X^\pm=X\pm g(X)\in \Gamma(C_\pm)$,
The notations are slightly changed from \cite{David2007}.
It satisfies the axioms of the connection, and preserves the $O(d,d)$-inner product.

The generalized torsion $\tau$ and the curvature $\Omega$ of $D$ are defined by 
\bea
	&{\tau}(X,Y)=D_X Y^+ - D_Y X^+ - ([X,Y])^+,\nn
	&{\Omega}(X,Y)u=(D_X D_Y - D_YD_X - D_{[X,Y]})u.
	\label{riemtenstand0}
\ena
They are tensors of $
\Gamma(\wedge^2 T^*M \otimes C_+)$ and
$
\Gamma(\wedge^2 T^*M \otimes {\rm End}(C_+))$, respectively.

In local coordinates $\{ x^i\}$ with the basis $(\p_j)^+$ of $C_+$,
the connection coefficient reads
\bea
	D_{\p_i} (\p_j)^+ &= \Upsilon^{m}_{ij}  (\p_m)^+, ~~
	\Upsilon^m_{ij}=\Gamma^m_{ij}+\frac{1}{2}g^{mk} H_{kij}, \label{Hconnection}
\ena
where $\Gamma^m_{ij}$ is the Christoffel symbol of the Levi-Civita connection, 
that is constructed from the metric $g_{\mu\nu}$.
It shows that the generalized connection $D$ on $C_+$ is the lift of an affine connection on $TM$,
such that an $H$-flux appears as a deviation from the Levi-Civita connection.

The $H$-flux arises as the torsion tensor, as seen from the components of the torsion tensor, 
\bea
	\tau_{ij}^m =\Upsilon^m_{ij} -\Upsilon^m_{ji}=g^{mk} H_{kij}.
\ena
The components of the generalized curvature
\bea
	{\Omega}^m_{kij} = \p_i  \Upsilon^m_{jk}-\p_j \Upsilon^m_{ik}
	 + \Upsilon^l_{jk}  \Upsilon^m_{il} 
	 - \Upsilon^l_{ik}  \Upsilon^m_{jl},
\ena 
is written by using \eqref{Hconnection} as
\bea
	{\Omega}^m_{kij}
	   &={\sf R}^{m}_{kij} 
	    +  \frac{1}{4}g^{ln} H_{njk} g^{mp} H_{pil} 
	    -  \frac{1}{4}g^{ln} H_{nik} g^{mp} H_{pjl}
	    +\frac{1}{2} g^{ml} \nabla_i H_{ljk} -\frac{1}{2}g^{ml} \nabla_j H_{lik},
\ena
where ${\sf R}^m_{kij}$ is the  
Riemann curvature tensor
\bea
	{\sf R}^m_{kij}=\p_i  \Gamma^m_{jk}-\p_j  \Gamma^m_{ik}
	 + \Gamma^l_{jk}  \Gamma^m_{il} - \Gamma^l_{ik}  \Gamma^m_{jl}.
\ena
Then, the generalized Ricci tensor is obtained by taking a contraction:
\bea
	{\Omega}_{kj}
	&:={\Omega}^m_{kmj} 
	    ={{\sf R}_{kj}}
	    -  \frac{1}{4}g^{ln} H_{nmk} g^{mp} H_{pjl}
	    +\frac{1}{2} g^{ml} \nabla_m H_{ljk},
\ena
with the Ricci tensor ${\sf R}_{kj}$, and the generalized Ricci scalar is given by
\bea
	{\Omega}&:=
	g^{kj} {\Omega}_{kj} 
	={\sf R}
	    -  \frac{1}{4} {g^{il}g^{jm}g^{kn} H_{ijk}  H_{lmn}} , \label{generalizedRicci}
\ena
where ${\sf R}$ is the usual Ricci scalar constructed by the Riemann metric $g$.
This form \eqref{generalizedRicci} of the generalized Ricci scalar  
appears in the action of the NS-NS sector of the supergravity.

\section{Computational details \label{compdet}}

\paragraph{Proofs of metricity \eqref{28} and torsion-less condition \eqref{torsion-free condition}}

\begin{itemize}
\item{Proof of \eqref{28}:}
By the Koszul formula, we have
\bea
&2G (\nabla_\xi \eta, \zeta) +2G (\eta, \nabla_\xi \zeta) \nn
=&\theta(\xi)\cdot G(\eta,\zeta)+\theta(\eta)\cdot G(\xi,\zeta)-\theta(\zeta)\cdot G(\xi,\eta)
+\theta(\xi)\cdot G(\zeta,\eta)+\theta(\zeta)\cdot G(\xi,\eta)-\theta(\eta)\cdot G(\xi,\zeta) \nn
&~~+G([\zeta,\xi]_\theta,\eta)+G([\zeta,\eta]_\theta,\xi)+G([\xi,\eta]_\theta,\zeta)
+G([\eta,\xi]_\theta,\zeta)+G([\eta,\zeta]_\theta,\xi)+G([\xi,\zeta]_\theta,\eta)\nn
=&2\theta(\xi)\cdot G(\eta,\zeta)\nn
=&2\bar\cL_{\xi} G(\eta,\zeta),
\ena
which says that $\nabla$ is metric-compatible.
\item{Proof of \eqref{torsion-free condition}:}
By using the Koszul formula again, we have
\bea
&2G(\bar T(\xi,\eta),\zeta)\nn
=&2G (\nabla_\xi \eta, \zeta)-2G (\nabla_\eta \xi, \zeta)-2G ([\xi, \eta]_\theta, \zeta)\nn
=&\theta(\xi)\cdot G(\eta,\zeta)+\theta(\eta)\cdot G(\xi,\zeta)-\theta(\zeta)\cdot G(\xi,\eta)
-\theta(\eta)\cdot G(\xi,\zeta)-\theta(\xi)\cdot G(\eta,\zeta)+\theta(\zeta)\cdot G(\eta,\xi)\nn
&~~+G([\zeta,\xi]_\theta,\eta)+G([\zeta,\eta]_\theta,\xi)
+G([\xi,\eta]_\theta,\zeta) -G([\zeta,\eta]_\theta,\xi)-G([\zeta,\xi]_\theta,\eta)-G([\eta,\xi]_\theta,\zeta)\nn
&~~-2G ([\xi, \eta]_\theta, \zeta)\nn
=&0,
\ena
for arbitrary $\xi$, $\eta$ and $\zeta$, which shows that $\bar T=0$.
\end{itemize}

\paragraph{Proofs of the Bianchi intensities \eqref{bi1} and \eqref{bi2}}

For the curvature for a contravariant affine connection $\bar{\nabla}$,
 together with the torsion tensor, we have the Bianchi identities:
\begin{itemize}
\item{First Bianchi identity \eqref{bi1}:} Since $\bar{T}(\xi,\eta)=\bar{\nabla}_\xi\eta-\bar{\nabla}_\eta\xi-[\xi,\eta]_\theta$\footnote{
In the case where $\bar{\nabla}$ is the contravariant Levi-Civita, the torsion tensor vanishes, $\bar{T}=0$.},
we have
\bea
	\bar{\nabla}_{\zeta}[\bar{T}(\xi,\eta)]
	&=\bar{\nabla}_{\zeta}\bar{\nabla}_\xi\eta-\bar{\nabla}_{\zeta}\bar{\nabla}_\eta\xi
	-(\bar{T}(\zeta,[\xi,\eta]_\theta) +\bar{\nabla}_{[\xi,\eta]_\theta}\zeta +[\zeta, [\xi,\eta]_\theta]_\theta).
\ena
Then we find
\bea
	\bar{\nabla}_{\zeta}&[\bar{T}(\xi,\eta)]+
	\bar{\nabla}_{\xi}[\bar{T}(\eta,\zeta)]+
	\bar{\nabla}_{\eta}[\bar{T}(\zeta,\xi)]\nn
	=& \bar{R}(\zeta,\xi) \eta +\bar{R}(\xi,\eta) \zeta + \bar{R}(\eta,\zeta)\xi
	-(\bar{T}(\zeta,[\xi,\eta]_\theta)+\bar{T}(\xi,[\eta,\zeta]_\theta)+\bar{T}(\eta,[\zeta,\xi]_\theta)),
\ena
where we used the Jacobi identity 
$[\zeta, [\xi,\eta]_\theta]_\theta+[\xi, [\eta,\zeta]_\theta]_\theta+[\eta, [\zeta,\xi]_\theta]_\theta=0$.
On the other hand, since $\bar{T}$ is tensorial, we also have
\bea
	\bar{\nabla}_{\zeta}[\bar{T}(\xi,\eta)]
	&=\bar{\nabla}_{\zeta}\bar{T}(\xi,\eta)+\bar{T}(\bar{\nabla}_{\zeta}\xi,\eta)
	+\bar{T}(\xi,\bar{\nabla}_{\zeta}\eta).
\ena
Then we find the first Bianchi identity
\bea
	&\bar{R}(\zeta,\xi) \eta +\bar{R}(\xi,\eta) \zeta + \bar{R}(\eta,\zeta)\xi \nn
	&=\bar{\nabla}_{\zeta}\bar{T}(\xi,\eta)+\bar{\nabla}_{\xi}\bar{T}(\eta,\zeta)
	+\bar{\nabla}_{\eta}\bar{T}(\zeta,\xi)+\bar{T}(\bar{T}(\xi,\eta),\zeta)
	+\bar{T}(\bar{T}(\eta,\zeta),\xi)+\bar{T}(\bar{T}(\zeta,\xi),\eta).
\ena
\item{Second Bianchi identity \eqref{bi2}:} Noting
\bea
	\bar{\nabla}_\zeta [\bar{R}(\xi,\eta) \omega]
	&=\bar{\nabla}_\zeta (\bar{R}(\xi,\eta)) \omega
	+ \bar{R}(\xi,\eta) \bar{\nabla}_\zeta\omega \nn
	&=(\bar{\nabla}_\zeta\bar{\nabla}_\xi \bar{\nabla}_\eta -\bar{\nabla}_\zeta\bar{\nabla}_\eta\bar{\nabla}_\xi
	 -\bar{R}(\zeta,[\xi,\eta]_\theta) - \bar{\nabla}_{[\xi,\eta]_\theta} \bar{\nabla}_\zeta
	 - \bar{\nabla}_{[\zeta,[\xi,\eta]_\theta]_\theta}) \omega,
\ena
we obtain, with the use of the Jacobi identity,
\bea
	[\bar{\nabla}_\zeta (\bar{R}(\xi,\eta))
	+\bar{\nabla}_\xi (\bar{R}(\eta,\zeta)) 
	+\bar{\nabla}_\eta (\bar{R}(\zeta,\xi)) ]\omega
	=-[\bar{R}(\zeta,[\xi,\eta]_\theta)+\bar{R}(\xi,[\eta,\zeta]_\theta)+\bar{R}(\eta,[\zeta,\xi]_\theta)]\omega.
\ena	
Noticing that $\bar{\nabla}_\zeta (\bar{R}(\xi,\eta))\omega=\bar{\nabla}_\zeta \bar{R}(\xi,\eta)\omega
+\bar{R}(\bar{\nabla}_\zeta\xi,\eta)\omega+\bar{R}(\xi,\bar{\nabla}_\zeta\eta)\omega$,
as $\bar{R}$ is tensorial, we find the second Bianchi identity
\bea
	(\bar{\nabla}_\zeta \bar{R}(\xi,\eta)
	+\bar{\nabla}_\xi \bar{R}(\eta,\zeta) 
	+\bar{\nabla}_\eta \bar{R}(\zeta,\xi) )\omega
	=(\bar{R}(\zeta,\bar{T}(\xi,\eta))+\bar{R}(\xi,\bar{T}(\eta,\zeta))+\bar{R}(\eta,\bar{T}(\zeta,\xi)))\omega.
\ena
\end{itemize}

\paragraph{Transformation law for $\bar \Gamma^{ij}_k$ \eqref{trf law of bar Gamma}}
The contravariant derivative of a $1$-form is given by
\bea
\bar \nabla^i \eta_j =\p^i \eta_j +\bar \Gamma^{ik}_j \eta_k.
\ena
By a $\beta$-diffeomorphism, we have
\bea
\bar \nabla^{'\, i} \eta'_j =\p^i \eta'_j +\bar \Gamma^{'\, ik}_j \eta'_k.
\ena
where $\eta'_j= \eta_j +\bar {\cal L}_\zeta \eta_j $.
If we write $\bar \Gamma^{'\, ik}_j =\bar \Gamma^{ik}_j +\delta_\zeta \bar \Gamma^{ik}_j$,
then it reduces (at the linear order in $\zeta$)
\bea
\bar \nabla^{'\, i} \eta'_j 
&=\bar \nabla^i (\eta_j +\bar {\cal L}_\zeta \eta_j) +\delta_\zeta \bar \Gamma^{ik}_j \eta_k .
\ena
On the other hand, we demand that it transforms as a $(1,1)$-tensor.
That is, 
\bea
\bar \nabla^{'\, i} \eta'_j 
&=\bar \nabla^i \eta_j +\bar {\cal L}_\zeta (\bar \nabla^i \eta_j).
\ena
Then, $\delta_\zeta \bar \Gamma^{ik}_j$ should be determined by
\bea
\delta_\zeta \bar \Gamma^{ik}_j \eta_k
&=\bar {\cal L}_\zeta (\bar \nabla^i \eta_j) -\bar \nabla^i (\bar {\cal L}_\zeta \eta_j) .
\label{delta bar Gamma}
\ena
We now compute the right hand side.

From the transformation law of a $(1,1)$-tensor,
\bea
\bar {\cal L}_\zeta (\bar \nabla^i \eta_j)
&=\zeta_k \p^k (\bar \nabla^i \eta_j ) + M^k_j (\bar \nabla^i \eta_k) -M^i_k (\bar \nabla^k \eta_j) \nn
&=\zeta_k \p^k (\p^i \eta_j +\bar \Gamma^{il}_j \eta_l) 
+ M^k_j (\p^i \eta_k +\bar \Gamma^{il}_k \eta_l) 
-M^i_k (\p^k \eta_j +\bar \Gamma^{kl}_j \eta_l) \nn
&=\zeta_k \p^k \p^i \eta_j +\zeta_k (\p^k \bar \Gamma^{il}_j \eta_l+ \bar \Gamma^{il}_j  \p^k \eta_l )
+ M^k_j (\p^i \eta_k +\bar \Gamma^{il}_k \eta_l) 
-M^i_k (\p^k \eta_j +\bar \Gamma^{kl}_j \eta_l)
\ena
On the other hand, we have
\bea
\bar \nabla^i (\bar {\cal L}_\zeta \eta_j)
&=\p^i  (\bar {\cal L}_\zeta \eta_j) +\bar \Gamma^{ik}_j  (\bar {\cal L}_\zeta \eta_k) \nn
&=\p^i (\zeta_k \p^k \eta_j +\eta_k M^k_j ) 
+\bar \Gamma^{ik}_j (\zeta_l \p^l \eta_k +\eta_l M^l_k) \nn
&=\p^i \zeta_k \p^k \eta_j +\zeta_k \p^i\p^k \eta_j
+\p^i \eta_k M^k_j +\eta_k \p^i M^k_j 
+\bar \Gamma^{ik}_j (\zeta_l \p^l \eta_k +\eta_l M^l_k)
\ena
From these expressions, \eqref{delta bar Gamma} is written as
\bea
\delta_\zeta \bar \Gamma^{ik}_j \eta_k
&=\zeta_k (\p^k \p^i -\p^i \p^k) \eta_j  -\p^i \zeta_k \p^k \eta_j  -M^i_k \p^k \eta_j \nn
&~~+\left( \zeta_k \p^k \bar \Gamma^{il}_j
+M^k_j \bar \Gamma^{il}_k -M^i_k \bar \Gamma^{kl}_j -m^l_k \bar \Gamma^{ik}_j
-\p^i M^l_j \right)\eta_l ,
\label{pre trf law of bar Gamma}
\ena
but the first line vanishes (see below) and we finally obtain 
\bea
\delta_\zeta \bar \Gamma^{il}_j 
&=\zeta_k \p^k \bar \Gamma^{il}_j
+M^k_j \bar \Gamma^{il}_k -M^i_k \bar \Gamma^{kl}_j -M^l_k \bar \Gamma^{ik}_j
-\p^i M^l_j .
\label{trf law of bar Gamma2}
\ena
We show that the first line in \eqref{pre trf law of bar Gamma} vanishes.
By using $M^i_k=-(\p^i \zeta_k +\p_k \theta^{il}\zeta_l)$, we have
\bea
&\zeta_k (\p^k \p^i -\p^i \p^k) \eta_j  -\p^i \zeta_k \p^k \eta_j  -M^i_k \p^k \eta_j \nn
=&\zeta_l \left( (\p^l \p^i -\p^i \p^l) \eta_j + \p_k \theta^{il} \p^k \eta_j \right).
\ena
By noting that
\bea
(\p^l \p^i -\p^i \p^l) \eta_j 
&=\theta^{ln}\p_n (\theta^{im}\p_m \eta_j ) -\theta^{im}\p_m (\theta^{ln}\p_n \eta_j ) \nn
&=\theta^{ln}\p_n \theta^{im} \p_m \eta_j  -\theta^{im}\p_m \theta^{ln}\p_n \eta_j  \nn
&=(\theta^{ln}\p_n \theta^{im}-\theta^{in}\p_n \theta^{lm})\p_m \eta_j, 
\ena
we thus have
\bea
\zeta_l \left( (\p^l \p^i -\p^i \p^l) \eta_j + \p_k \theta^{il} \p^k \eta_j \right) 
=&\zeta_l (\theta^{ln}\p_n \theta^{im}-\theta^{in}\p_n \theta^{lm} 
+\p_n \theta^{il} \theta^{nm} )\p_m \eta_j \nn
=& \zeta_l (\theta^{ln}\p_n \theta^{im}+\theta^{in}\p_n \theta^{ml} 
+\theta^{mn}\p_n \theta^{li}  )\p_m \eta_j \nn
=&0,
\ena
due to the Poisson condition.

\paragraph{Contravariant derivative of a tensor field  \eqref{contra dedivative law}}
To show that claim,
$\bar{\nabla}_\zeta T^{i_1 \cdots i_r}_{j_1 \cdots j_s}$ transforms as a $(r,s)$-tensor again,
define the difference as ${\cal A}_\zeta =\bar{\cal L}_\zeta -\bar{\nabla}_\zeta$, where 
$\bar{\nabla}_\zeta$ is a affine contravariant derivative.
Then, by using and (\ref{beta tensor law}) and (\ref{contra dedivative law}), we have on an $(r,s)$-tensor $T$, 
\bea
({\cal A}_\zeta T)^{i_1 \cdots i_r}_{j_1 \cdots j_s}
&=\sum_{p=1}^r \left( -M^{i_p}_l 
+\bar{\Gamma}^{ki_p}_l \zeta_k \right)
T^{i_1 \cdots l \cdots i_r}_{j_1 \cdots j_s} -\sum_{q=1}^s 
\left( -M^l_{j_q}+\bar{\Gamma}^{kl}_{j_q} \zeta_k \right)
 T^{i_1 \cdots i_r}_{j_1 \cdots l \cdots j_s}.
\ena
By using (\ref{contra 1-form law}) 
$\bar{\nabla}^l \zeta_{j_q} 
=\p^l \zeta_{j_q}  
+\bar{\Gamma}^{lk}_{j_q} \zeta_k$,
the term inside the bracket becomes 
\bea
-M^l_{j_q}+\bar{\Gamma}^{kl}_{j_q} \zeta_k
&=\p^{l} \zeta_{j_q} +\p_{j_q} \theta^{lk} \zeta_k 
+\bar{\Gamma}^{kl}_{j_q} \zeta_k \nn
&=\bar{\nabla}^{l} \zeta_{j_q} +\p_{j_q} \theta^{lk} \zeta_k 
+\left(\bar{\Gamma}^{kl}_{j_q} 
-\bar{\Gamma}^{lk}_{j_q} \right)\zeta_k\nn
&=\bar{\nabla}^{l} \zeta_{j_q} -T^{kl}_{j_q}\zeta_k.
\ena
Therefore, we can simplify the result as
\bea
({\cal A}_\zeta T)^{i_1 \cdots i_r}_{j_1 \cdots j_s}
&=\sum_{p=1}^r (\bar{\nabla}^{i_p} \zeta_{l} -T^{ki_p}_l\zeta_k)
T^{i_1 \cdots l \cdots i_r}_{j_1 \cdots j_s}
-\sum_{q=1}^s (\bar{\nabla}^{l} \zeta_{j_q} -T^{kl}_{j_q}\zeta_k)
 T^{i_1 \cdots i_r}_{j_1 \cdots l \cdots j_s}.
\ena
Since $\bar{\nabla}^l \zeta_{j}$ or equivalently $\bar{\nabla}_\xi \zeta$ is a tensor, 
this expression is a combination of tensor fields only. Thus, the claim is shown.

\paragraph{Proof of (\ref{riempois}) \label{sriempois}} 
By using \eqref{covariant-contravariant}, we compute preliminarily
\bea
	\theta^{im} \p_m {\bar{\Gamma}}^{jk}_l
		 &= \theta^{im} 
		  (\p_m\Gamma^k_{nl} \theta^{nj} + \Gamma^k_{nl}\p_m \theta^{nj}    
		+ \p_m K^{jk}_l ), \\
	(\p_n \theta^{ij}) {\bar{\Gamma}}^{nk}_l 
		&= (\p_n \theta^{ij}) 
		(  \Gamma^k_{pl} \theta^{pn}    
		+K^{nk}_l ),\\
	{\bar{\Gamma}}^{jk}_m {\bar{\Gamma}}^{im}_l  
	&= 
		(  \Gamma^k_{nm} \theta^{nj}    
		+K^{jk}_m  )
		(  \Gamma^m_{pl} \theta^{pi}    
		+K^{im}_l  ).
\ena
Gathering these stuff, we find that the contravariant Riemann tensor in \eqref{contra Riem} results in
\bea
	\bar{R}^{kij}_{l}
	&= \theta^{im}  (\p_m\Gamma^k_{nl}) \theta^{nj} 
		- \theta^{jm} (\p_m\Gamma^k_{nl}) \theta^{ni} 
		+ \Gamma^k_{nm} \theta^{nj}  \Gamma^m_{pl} \theta^{pi} 
		- \Gamma^k_{nm} \theta^{ni}  \Gamma^m_{pl} \theta^{pj}    \label{b4}\\
	&	
		+\theta^{im} \p_m K^{jk}_l 
		-\theta^{jm}  \p_m K^{ik}_l 
		+\Gamma^k_{nm} \theta^{nj}  K^{im}_l  
		-  K^{ik}_m \Gamma^m_{pl} \theta^{pj} 
		+ K^{jk}_m \Gamma^m_{pl} \theta^{pi}  
		-  \Gamma^k_{nm} \theta^{ni} K^{jm}_l 
		 \label{b5}\\
	&
	+ \theta^{im}\Gamma^k_{nl}\p_m \theta^{nj}    
		- \theta^{jm} \Gamma^k_{nl}\p_m \theta^{ni}   
		+K^{jk}_m  K^{im}_l 
		- K^{ik}_mK^{jm}_l 
		- (\p_n \theta^{ij}) 
		(  \Gamma^k_{pl} \theta^{pn}    
			+K^{nk}_l ) .	\label{Riem}
\ena
The first line \eqref{b4} is resulted in the celebrated Riemann curvature tensor:
\bea
	\eqref{b4}
	&=\theta^{im} \theta^{nj}  (\p_m\Gamma^k_{ln}
		- \p_n\Gamma^k_{lm}
		+ \Gamma^p_{ln} \Gamma^k_{pm}   
		-    \Gamma^p_{lm}\Gamma^k_{pn} 
		 )  = \theta^{im} \theta^{nj} {\sf R}^k_{lmn},
\ena
where in our notation 
it is defined by 
\bea
	{\sf R}^k_{lmn} = \p_m\Gamma^k_{ln}
		- \p_n\Gamma^k_{lm}
		+ \Gamma^p_{ln} \Gamma^k_{pm}   
		-    \Gamma^p_{lm}\Gamma^k_{pn},
\ena
as usual.
The second line \eqref{b5} is rewritten in terms of covariant derivatives as
\bea
	\eqref{b5}=  \theta^{nj} \nabla_n K^{ik}_l -\theta^{nj}\Gamma^i_{nm}   K^{mk}_l  
		-\theta^{ni} \nabla_n K^{jk}_l +\theta^{ni}\Gamma^j_{nm}   K^{mk}_l .
\ena
Hence, so far the contravariant curvature is summarized as
\bea
	\bar{R}^{kij}_{l}
	&= \theta^{im} \theta^{nj} {\sf R}^k_{lmn}
 		+ \theta^{nj} \nabla_n K^{ik}_l 
		 -\theta^{ni} \nabla_n K^{jk}_l
		 +K^{jk}_m  K^{im}_l 
		- K^{ik}_mK^{jm}_l
		  \nn
	&    
		 + \theta^{im}\Gamma^k_{nl}\p_m \theta^{nj}    
		- \theta^{jm} \Gamma^k_{nl}\p_m \theta^{ni}   
		- (\p_n \theta^{ij}) 
		(  \Gamma^k_{pl} \theta^{pn}    
			+K^{nk}_l )
			- \theta^{nj}\Gamma^i_{nm}   K^{mk}_l  
		 +\theta^{ni}\Gamma^j_{nm}   K^{mk}_l.	\label{Riema}
\ena
The second line in this expression 
still has terms involving the partial derivatives and the Christoffel  symbols,
but it turns out to be reduced to a covariant tensor
\bea
	(\text{second line of \eqref{Riema}})=-  (\p_n \theta^{ij}  + \Gamma^i_{mn}   \theta^{mj}  
		 -\Gamma^j_{mn} \theta^{mi}   )K^{nk}_l 
		 = -  (\nabla_n \theta^{ij}) K^{nk}_l.
\ena
Here we used \eqref{contra Poisson condition}, which is equivalent to the Poisson condition.
Finally, we obtain the result 
\bea
	\bar{R}^{kij}_{l}
	&= \theta^{im} \theta^{nj} {\sf R}^k_{lmn}
		-  (\nabla_n \theta^{ij}) K^{nk}_l
 		+ \theta^{nj} \nabla_n K^{ik}_l 
		 -\theta^{ni} \nabla_n K^{jk}_l
		+K^{jk}_m  K^{im}_l 
		- K^{ik}_mK^{jm}_l  . 
\ena

\paragraph
{Proofs of  (\ref{bracketproperty1}) and (\ref{bracketproperty2}) \label{bracketproperty}}

In this paragraph,  we denote
$f,g \in C^\infty(M)$ 
and $u=X+\xi$, $v=Y+\eta$, $w=Z+\zeta \in \Gamma(TM\oplus T^*M)$.
To show the desired relations, we preliminarily list  the formulae following by definitions and from some
manipulations:
\bea
	&\bar{\cL}_{(f\xi)} g = f \bar{\cL}_\xi g,	\label{A.1}\\
	&\bar{\cL}_{(f\xi)} \zeta 
	= f \bar{\cL}_{\xi} \zeta -  (\bar{\cL}_{\zeta} f) \xi,	\label{A.2}\\
	&\bar{\cL}_{(f\xi)} X
	= f \bar{\cL}_\xi X + ( \bar{\iota}_{\xi}X ) d_\theta f,\\
	&\bar{\cL}_\xi (fg) 
	= (\bar{\cL}_\xi f)g +  f\bar{\cL}_\xi g  ,\\
	&\bar{\cL}_\xi (f\zeta)
	=  f \bar{\cL}_{\xi} \zeta +  (\bar{\cL}_{\xi} f) \zeta,	\label{A.5}\\
	&\bar{\cL}_\xi (fX)
	=  f\bar{\cL}_\xi X  + (\bar{\cL}_\xi f) X,
\ena
where we used $ d_\theta  (fX)=  d_\theta f\wedge X + fd_\theta X$.
With these preliminaries, first, (\ref{bracketproperty1}) is shown by
\bea
	&[X+\xi,f(Y+\eta)] 
	= \bar{\cL}_{\xi}[ f(Y+\eta) ]- \bar{\cL}_{(f\eta)} X
	+ \frac{1}{2} d_\theta(\bar{\iota}_{(f\eta)}X - \bar{\iota}_\xi {(fY)})  \nn
	&= f\bar{\cL}_{\xi} (Y+\eta) + (\bar{\cL}_\xi f) (Y+\eta)
		- f \bar{\cL}_{\eta} X - (\bar{\iota}_\eta X) d_\theta f
		+ \frac{1}{2}f d_\theta(\bar{\iota}_\eta X - \bar{\iota}_\xi{Y})  
		+ \frac{1}{2} (d_\theta f) (\bar{\iota}_\eta X- \bar{\iota}_\xi Y)  \nn
	&=f[X+\xi,Y+\eta] 	 + (\bar{\cL}_\xi f) (Y+\eta) -  (d_\theta f) \langle X+\xi,Y+\eta \rangle.
\ena
Second, the equation (\ref{bracketproperty2}) is shown by
\bea
	&\langle [u,v]+d_\theta \langle u,v\rangle, w\rangle
	+ \langle v, [ u,w]+d_\theta \langle u, w\rangle \rangle \nn
	&=\langle \bar{\cL}_\xi v- \bar{\iota}_\eta d_\theta X - d_\theta \bar{\iota}_\eta X 
	+ d_\theta  \bar{\iota}_\eta X , w\rangle
	+ \langle v, [ u,w]+d_\theta \langle u, w\rangle \rangle \nn
	&=\langle \bar{\cL}_\xi (Y+\eta)- \bar{\iota}_\eta d_\theta X   , w\rangle
	+ \langle v, [ u,w]+d_\theta \langle u, w\rangle \rangle \nn 
	&=\frac{1}{2}( \bar{\iota}_\zeta \bar{\cL}_\xi Y
		+  \bar{\iota}\bar{\cL}_\xi \eta- \bar{\iota}_\zeta \bar{\iota}_\eta d_\theta X   )
	+\frac{1}{2}( \bar{\iota}_\eta \bar{\cL}_\xi Z
		+  \bar{\iota}_Y \bar{\cL}_\xi \zeta- \bar{\iota}_\eta \bar{\iota}_\zeta d_\theta X   ) 
	=\frac{1}{2}\bar{\cL}_\xi  (\bar{\iota}_\zeta Y + \bar{\iota}_Z \eta) 
	= \bar{\cL}_\xi \langle v,w \rangle ,
\ena
where we used $ \bar{\iota}_\zeta \bar{\cL}_\xi Y + i_Y \bar{\cL}_\xi \zeta  = \bar{\cL}_\xi (\bar{\iota}_\zeta Y) .$

\paragraph{Proofs of  (\ref{connection1}) and (\ref{connection2}) \label{connection}}

We denote $\xi \in \Gamma(T^*M)$, $\xi^-=\xi-G(\xi)\in \Gamma(C_-)$, $u=X+\zeta \in \Gamma(C_+)$. 
By definition and (\ref{bracketproperty1}),  
we obtain
\bea
	\bar{D}_{(f\xi)} u 
	&= \pi_+ ([f\xi^-,u]_{R}) \nn
	&=\pi_+ (f[\xi^-,u]_{R}-(\bar{\cL}_\zeta f)\xi^- + d_\theta f \langle \xi^- ,u\rangle ) 
	= \pi_+ (f[\xi^-,u] ) 
	= f\bar{D}_{\xi} u,
\ena
where we used $\pi_+(\xi^-)=0$ and $ \langle \xi^- ,u\rangle =0$. Similarly, we find
\bea
	\bar{D}_{\xi} (fu) 
	&= \pi_+ ([\xi^-,fu]_{R}) \nn
	&=\pi_+ (f[\xi^-,u]_{R}+(\bar{\cL}_\xi f)u - d_\theta f \langle \xi^- ,u\rangle ) \nn
	&= \pi_+ (f[\xi^-,u]_{R} )+(\bar{\cL}_\xi f)u 
	= f\bar{D}_{\xi} u +(\bar{\cL}_\xi f)u,
\ena
where $\pi_+(u)=u$ is used. 

\paragraph{Proof of (\ref{compatible}) \label{compatibility}}

By using (\ref{bracketproperty2}) for $u=\xi^-$ (note the footnote after (\ref{bracketproperty2})), we have
\bea
	\bar{\cL}_\xi \langle u,v \rangle
	&=  \langle [\xi^-,u]+d_\theta  \langle \xi^- , u  \rangle  , v  \rangle  
		+  \langle u , [\xi^-,v]  +d_\theta  \langle \xi^- , v  \rangle\rangle \nn
	&=  \langle [\xi^-,u]  , v  \rangle  
		+  \langle u , [\xi^-,v]  \rangle \nn
	&=  \langle  \pi_+[\xi^-,u]  , v  \rangle  
		+  \langle u , \pi_+ [\xi^-,v]  \rangle 
	=  \langle \bar{D}_\xi u  , v  \rangle  
		+  \langle u ,  \bar{D}_\xi v  \rangle	.
\ena

\paragraph{Proof of (\ref{CurvatureTensor}) \label{CT}}

First, by combining the results \eqref{A.1}, \eqref{A.2} and \eqref{A.5}, we find
\bea
	[f\xi, g\eta]&=[f\xi, g\eta]_\theta \nn
	&=f\bar{\cL}_\xi(g\eta)-(\bar{\cL}_{(g\eta)}f)\xi \nn
	&=fg\bar{\cL}_\xi \eta+ f(\bar{\cL}_\xi g)\eta-g(\bar{\cL}_{\eta}f)\xi 
	=fg [\xi, \eta]+ f(\bar{\cL}_\xi g)\eta-g(\bar{\cL}_{\eta}f)\xi .	\label{a20}
\ena
Second, from \eqref{connection1} and \eqref{connection2} we obtain
\bea
	\bar{D}_{f\xi} \bar{D}_{g\eta} (hu)
	&=f \bar{D}_\xi (gh\bar{D}_\eta u + (g \bar{\cL}_\eta h)u) \nn
	&=f  (\bar{\cL}_\xi(gh)\bar{D}_\eta u   +gh \bar{D}_\xi \bar{D}_\eta u+ \bar{\cL}_\xi(g \bar{\cL}_\eta h)u
		+(g \bar{\cL}_\eta h)\bar{D}_\xi u),
\ena
with $u=\zeta^+$.
Hence, on one hand, we find that
\bea
	&(\bar{D}_{f\xi} \bar{D}_{g\eta}-\bar{D}_{g\eta}\bar{D}_{f\xi}  )(hu) \nn
	&=f  (\bar{\cL}_\xi(gh)\bar{D}_\eta u   +gh \bar{D}_\xi \bar{D}_\eta u+ \bar{\cL}_\xi(g \bar{\cL}_\eta h)u
		+(g \bar{\cL}_\eta h)\bar{D}_\xi u) \nn
		&~~~ - g  (\bar{\cL}_\eta(fh)\bar{D}_\xi u   +fh \bar{D}_\eta \bar{D}_\xi u+ \bar{\cL}_\eta(f \bar{\cL}_\xi h)u
		+(f \bar{\cL}_\xi h)\bar{D}_\eta u) \nn
	&= fgh (\bar{D}_\xi \bar{D}_\eta - \bar{D}_\eta \bar{D}_\xi)u
		+fh (\bar{\cL}_\xi g) \bar{D}_\eta u - gh (\bar{\cL}_\eta f) \bar{D}_\xi u + f \bar{\cL}_\xi(g \bar{\cL}_\eta h)u
		- g  \bar{\cL}_\eta(f \bar{\cL}_\xi h)u,
\ena
and on the other hand,
\bea
	&\bar{D}_{[f\xi,g\eta]}hu 
	=\bar{D}_{fg [\xi, \eta]+ f(\bar{\cL}_\xi g)\eta-g(\bar{\cL}_{\eta}f)\xi} hu \nn
	&= (fg\bar{D}_{ [\xi, \eta]}+ f(\bar{\cL}_\xi g)\bar{D}_{\eta}- g(\bar{\cL}_{\eta}f)\bar{D}_{\xi}) hu \nn
	&= fgh\bar{D}_{ [\xi, \eta]}u 
	+ fh(\bar{\cL}_\xi g)\bar{D}_{\eta} u  - gh(\bar{\cL}_{\eta}f)\bar{D}_{\xi}u
	+ f(\bar{\cL}_\xi g)(\bar{\cL}_{\eta}f) u
	  - g(\bar{\cL}_{\eta}f)(\bar{\cL}_{\xi}h) u +fg (\bar{\cL}_{[\xi,\eta]}h) u .
\ena
This shows
\bea
	(\bar{D}_{f\xi} \bar{D}_{g\eta}-\bar{D}_{g\eta}\bar{D}_{f\xi} -\bar{D}_{[f\xi,g\eta]} )(hu)
	= fgh  (\bar{D}_{\xi} \bar{D}_{\eta}-\bar{D}_{\eta}\bar{D}_{\xi} -\bar{D}_{[\xi,\eta]} )u.
\ena

\paragraph{Proof of (\ref{bracketbeta0}) \label{sbracketbeta0}}

The bracket in (\ref{bracketbeta0}) 
is computed by \eqref{Sasa bracket} as follows:
\bea
	[dx^i -G^{ik}\p_k, dx^j+G^{jl}\p_l]  &= [dx^i , dx^j]_\theta+ i_{dx^i}d_\theta  (G^{jl}\p_l)
	+i_{dx^j} d_\theta  (G^{jl}\p_l) 
	+ d_\theta G^{ij}.
\ena
Each term results in
\bea
	[dx^i,dx^j]_\theta &=\p_k \theta^{ij}dx^k, \\
	i_{dx^i}d_\theta  (G^{jl}\p_l) 
	&= i_{dx^i}[ \frac{1}{2}\theta^{mn}\p_m\wedge \p_n, G^{jl}\p_l]_S \nn
	&=i_{dx^i}\bigg([ \frac{1}{2}\theta^{mn}\p_m, G^{jl}\p_l]_S\wedge \p_n
		-  [\p_n, G^{jl}\p_l]_S\wedge \frac{1}{2}\theta^{mn}\p_m \bigg) \nn
	&=  \theta^{mn}(\p_m G^{ji})  \p_n
		-\theta^{mi}(\p_m G^{jl}) \p_l -  G^{jl} ( \p_l \theta^{in}) \p_n ,\\
	i_{dx^j}d_\theta  (G^{il}\p_l) 
	&=  \theta^{mn}(\p_m G^{ij})  \p_n
		-\theta^{mj}(\p_m G^{il}) \p_l -  G^{il} ( \p_l \theta^{jn}) \p_n ,\\
	d_\theta G^{ij} &= - \theta(dG^{ij}) 
	= -\theta^{kl}(\p_kG^{ij} )\p_l.
\ena
Hence the bracket reads
\bea
	&[dx^i -G^{ik}\p_k, dx^j+G^{jl}\p_l]  \nn
	&= \p_k \theta^{ij}dx^k 
		+ [\theta^{mn}(\p_m G^{ji})  
		-\theta^{mi}(\p_m G^{jn})  -\theta^{mj}(\p_m G^{in})
		 -  G^{jl} ( \p_l \theta^{in})  
		 -  G^{il} ( \p_l \theta^{jn})] \p_n.
\ena

\paragraph{Proof of \eqref{gene Riem}} 
\bea
4(\bar{\Omega }^{kij}_{l}-\bar{R}^{kij}_{l}) 
&=2\theta^{im} \p_m ({R}^{jkn}G_{nl})
- 2\theta^{jm} \p_m ({R}^{ikn}G_{nl} )
-2\p_n \theta^{ij} {R}^{nkm}G_{ml} \nn
&+2\bar{\Gamma}^{jk}_m  {R}^{imn}G_{nl}
+2{R}^{jkn}G_{nm} \bar{\Gamma}^{im}_l
+{R}^{jkn}G_{nm}{R}^{imn'}G_{n'l} \nn
&-2\bar{\Gamma}^{ik}_m  {R}^{jmn}G_{nl}
-2{R}^{ikn}G_{nm} \bar{\Gamma}^{jm}_l
-{R}^{ikn}G_{nm}{R}^{jmn'}G_{n'l}\nn
&=2(\bar \nabla_{dx^i} {R}^{jkn}) G_{nl} - 2(\bar \nabla_{dx^j} {R}^{ikn})G_{nl} 
-2\p_n \theta^{ij} {R}^{nkm}G_{ml} \nn
&+2\bar{\Gamma}^{ij}_m R^{mkn}G_{nl} -2\bar{\Gamma}^{ji}_m R^{mkn}G_{nl}
+{R}^{jkn}G_{nm}{R}^{imn'}G_{n'l} -{R}^{ikn}G_{nm}{R}^{jmn'}G_{n'l}\nn
&=2(\bar \nabla_{dx^i} {R}^{jkn}) G_{nl} - 2(\bar \nabla_{dx^j} {R}^{ikn})G_{nl} 
-2\p_n \theta^{ij} {R}^{nkm}G_{ml} \nn
&+2\p_m \theta^{ij}R^{mkn}G_{nl} 
+{R}^{jkn}G_{nm}{R}^{imn'}G_{n'l} -{R}^{ikn}G_{nm}{R}^{jmn'}G_{n'l}\nn
&=2(\bar \nabla_{dx^i} {R}^{jkn}) G_{nl} - 2(\bar \nabla_{dx^j} {R}^{ikn})G_{nl} 
+{R}^{jkn}G_{nm}{R}^{imn'}G_{n'l} -{R}^{ikn}G_{nm}{R}^{jmn'}G_{n'l}\nonumber
\ena
where we used
\bea
\theta^{im} \p_m ({R}^{jkn}G_{nl}) 
&=\bar \nabla_{dx^i} ({R}^{jkn}G_{nl}) 
+\bar{\Gamma}^{ij}_m R^{mkn}G_{nl}+\bar{\Gamma}^{ik}_m R^{jmn}G_{nl}
-\bar{\Gamma}^{im}_l R^{jkn}G_{nm},
\ena
and $\bar \Gamma^{ij}_m -\bar \Gamma^{ji}_m =\p_m \theta^{ij}$.


\providecommand{\href}[2]{#2}\begingroup\raggedright
\endgroup

\end{document}